%% file: Main_v2.tex
\documentclass[superscriptaddress,amsmath,reprint,amssymb,aps,nofootinbib,pra,floatfix,longbibliography,showkeys]{revtex4-2}
\usepackage[T1]{fontenc}
\usepackage{graphicx}
\usepackage{amsmath,amssymb,amsfonts}
\usepackage{times}
\usepackage{bbold}
\usepackage{textcomp,gensymb}
\usepackage{epsfig}
\usepackage{xcolor}
\usepackage{fancyhdr}
\usepackage{booktabs}
\usepackage[colorlinks,citecolor=blue,linkcolor=blue,urlcolor=blue]{hyperref}
\usepackage{stmaryrd}
\SetSymbolFont{stmry}{bold}{U}{stmry}{m}{n}
\definecolor{red}{rgb}{0,0,0} 
\definecolor{bblue}{rgb}{0.1, 0.1, 0.6}
\definecolor{green}{rgb}{0.2,0.53,0.5}

\newcommand{\ket}[1]{|#1\rangle}
\newcommand{\bra}[1]{\langle#1|}

\newcommand{\ensavg}[1]{\langle#1\rangle}

\newcommand*{\Scale}[2][4]{\scalebox{#1}{$#2$}}%

\begin{document}
\pagestyle{plain}
\hypersetup{pageanchor=false}

\title{Exploiting Spatial Diversity in Earth-to-Satellite Quantum-Classical Communications}

\author{Ziqing Wang}\email{ziqing.wang1@unsw.edu.au}
\affiliation{School of Electrical Engineering and Telecommunications, University of New South Wales, Sydney, NSW 2052, Australia}

\author{Timothy C. Ralph}\email{ralph@physics.uq.edu.au}
\affiliation{Centre for Quantum Computation and Communication Technology, School of Mathematics and Physics, University of Queensland, St Lucia, QLD 4072, Australia}

\author{Ryan~Aguinaldo}\email{ryan.aguinaldo@ngc.com}
\affiliation{Northrop Grumman Corporation, San Diego, CA 92128, USA}

\author{Robert~Malaney}\email{r.malaney@unsw.edu.au}
\affiliation{School of Electrical Engineering and Telecommunications, University of New South Wales, Sydney, NSW 2052, Australia}

\date{\today}

\begin{abstract}
Despite being an integral part of the vision of quantum Internet, Earth-to-satellite (uplink) quantum communications have been considered more challenging than their satellite-to-Earth (downlink) counterparts due to the severe channel-loss fluctuations (fading) induced by atmospheric turbulence. 
The question of how to address the negative impact of fading on Earth-to-satellite quantum communications remains largely an open issue.
In this work, we explore the feasibility of exploiting spatial diversity as a means of fading mitigation in Earth-to-satellite Continuous-Variable (CV) quantum-classical optical communications. 
We demonstrate, via both our theoretical analyses of quantum-state evolution and our detailed numerical simulations of uplink optical channels, that the use of spatial diversity can improve the effectiveness of entanglement distribution through the use of multiple transmitting ground stations and a single satellite with multiple receiving apertures. We further show that the transfer of both large (classically-encoded) and small (quantum-modulated) coherent states can benefit from the use of diversity over fading channels.
Our work represents the first quantitative investigation into the use of spatial diversity for satellite-based quantum communications in the uplink direction, showing under what circumstances this fading-mitigation paradigm, which has been widely adopted in classical communications, can be helpful within the context of Earth-to-satellite CV quantum communications.

\end{abstract}


\maketitle

\section{Introduction}

Quantum communications via Low-Earth-Orbit (LEO) satellites have become a key enabler in extending, to a global scale, the unprecedented information security guaranteed by quantum physics. 
As an alternative to the Discrete Variable (DV) paradigm (which maps information to the discrete features of single photons), the Continuous-Variable (CV) paradigm (which encodes information onto the quadrature variables of optical fields) has been much touted as a promising candidate towards the practical implementation of quantum communications since CV quantum states can be readily generated and measured with high-speed and high-efficiency optical devices~\cite{PredictiveOutlook2019}.
Within the CV paradigm, the recently proposed concept of Simultaneous Quantum-Classical communication (SQCC)~\cite{SQCC2016,SQCC2018} represents a promising direction towards the simpler integration of global quantum and classical communication networks. 
Utilizing a form of quantum-classical signaling achieved by applying a small quantum modulation onto a large (classically-modulated) coherent state, SQCC allows quantum communications and classical communications (including those supporting quantum communications) to be implemented using the same communication infrastructure (e.g., the same set of transmitter and receiver) and the same optical pulse with minimum additional complexity.
In addition to proof-of-principle experimental demonstrations via optical fiber (e.g.,~\cite{SQCC_25kmFiber}), the feasibility of SQCC has been recently explored within the context of satellite-based quantum communications~\cite{SQCC_Matt2024}.  SQCC has the ability to embed multiple quantum protocols
such as Quantum Key Distribution (QKD)~\cite{GG02}, and Quantum Direct Communication (QDC)~\cite{CV-SQDC}, within a wider class of classical communication protocols.

Despite this exciting outlook for global quantum communications, it is known that the random refractive-index fluctuations within the Earth's atmosphere (commonly referred to as \textit{atmospheric turbulence}) can significantly hinder the feasibility of all types of Free-Space Optical (FSO) communications by introducing a fluctuating loss (usually referred to as \textit{fading}) via a plethora of effects, such as beam broadening, beam wandering, and beam-shape deformation~\cite{book}. 
Although satellite-Earth quantum communications are two-way, the satellite-to-Earth  downlink  has been predominantly considered in most studies due the fact that it suffers minimal fading due to the beam width at atmospheric entry being much larger than any turbulent eddy (see discussions in, e.g.,~\cite{UplinkIntensity,UpDown2009,SatElptBeam2019}). Indeed, both the recent demonstration of inter-continental quantum communication~\cite{sat-based_QNet} and the world's first large-scale quantum communication network~\cite{QKD_Global_2021} utilized downlink channels exclusively.

Despite the predominant use of downlink channels, the construction of a global-scale quantum network also requires the use of uplink (i.e., Earth-to-satellite) channels. Depending on the specific use-case scenario, uplink channels are sometimes preferred due to their unique advantages, especially when the limitation of resources at a satellite and the flexibility of a ground-station system are taken into consideration (see discussions in, e.g.,~\cite{UplinkQKD_ProofOfPrinciple2017,SatMedLink,MiciusRMP2022}). 
{{Indeed, the uplink configuration avoids deploying high-quality CV quantum entanglement sources on satellites, which is, at the moment, still impractical from a system complexity point of view. 
From an operational point of view, the uplink configuration benefits from deploying quantum light sources at the ground stations, bringing great flexibility and allowing for future upgrades. 
In principle, a satellite-based system adopting the uplink configuration can support various quantum information protocols using different quantum light sources that can be readily diagnosed, maintained, reconfigured, or switched out at the ground stations.
The satellite payload can also be more versatile (relative to one containing the quantum light source) since it may naturally be compatible with various quantum information protocols, including those requiring the satellite to play the simple role of a relay. 
Furthermore, although some well-known quantum information protocols can be implemented using solely downlink channels, the uplink configuration makes attacks directed at the receiver significantly more difficult in practice and is naturally required by many advanced quantum information protocols (a more detailed discussion regarding the usefulness of the uplink configuration in quantum communications is provided in Sec.~\ref{Sec:UplinkUsefulness} of~\cite{SM}).}} 
However, it is well-known that quantum-state transfer over an uplink channel is more challenging since the optical beam encounters atmospheric turbulence at the very start of its propagation with a beam width smaller than the outer scale of the turbulence (i.e. the upper cut off in the size of turbulent eddies)  at the transmission point.
The early-stage wavefront distortions imposed on the transmitted optical beam are exacerbated by the remaining transmission path to the satellite receiver, leading to significant beam wandering and beam broadening in the far field (see discussions in, e.g.,~\cite{book,UplinkIntensity,SatMedLink,UplinkIssue2020,EduardoEnhanced2021,MiciusRMP2022}). 
As a result, severe fading that significantly hinders the feasibility of FSO communications can be expected within an uplink channel. In extreme cases, the wandering beam may completely miss the small receiving aperture at the satellite, and the resulting ``deep fades'' can render FSO communications non-viable over an uplink channel. 

In FSO communications, despite the absence of multipath propagation, spatial diversity can still be introduced by adopting multiple transmitters and/or receivers since two subchannels spatially separated by a distance much larger than the atmospheric coherence length (usually on the order of centimeters) can be considered independent 
(see discussions in, e.g.,~\cite{OpticalDiversity2004} for general FSO channels and~\cite{FSOC_Review_VC,OCS_Review2017} for satellite-based FSO channels).
Indeed, previous works have suggested the usefulness of utilizing spatial diversity in terrestrial (e.g.,~\cite{OpticalRxDiversity1996,FSO_Diversity2002,OpticalMIMO2009,DiversityFSO_IncoherentCoherent2009,DiversityCoherentFSO2009,OAM_Diversity2016}), satellite-to-Earth (e.g.,~\cite{DiversitySatOptCommsGammaGamma2019,DiversitySatOptCommsGammaGamma2015,DiversitySatOptCommsCoherentDPSK2020,UplinkDownlinkDiversity2022}), as well as Earth-to-satellite (e.g.,~\cite{UplinkTxDiversity2010,UplinkTransmitD2016,UplinkDownlinkDiversity2022}) FSO  classical communications. 
The use of diversity has rarely been investigated within the context of \textit{quantum} communications. The first step in this direction has been carried out recently in the work~\cite{TCommsDiversityQuantumComms2020} which investigated the usefulness of spatial diversity in the discrimination of two small coherent states in an analytical model of the fading channel.
However, it can be argued that the discrimination of two small coherent states in~\cite{TCommsDiversityQuantumComms2020} has limited direct application to deployed CV-QKD.
More importantly, consideration of diversity usage for the purpose of quantum entanglement distribution through {{realistic Earth-to-satellite atmospheric fading channels}} remains unexplored -- a critical missing element in the construction of a global quantum internet. Also missing is a detailed study of {{diversity-assisted coherent-state transfer over fading channels within the context of SQCC.}} In this work we deliver on both of these missing elements.

Our novel findings are summarized as follows. We find that it is indeed feasible to exploit spatial diversity for performance enhancement in Earth-to-satellite quantum communications.
Specifically, our results show that the use of diversity can enhance the lower bound of the distributed entanglement, improving the performance of Gaussian quantum information protocols (e.g., standard CV-QKD) in the uplink direction.
Within the context of SQCC over fading channels, our further results indicate that the transfer of both large (classically-modulated) coherent states (for the classical part of communications) and small (quantum-modulated) coherent states (for the quantum part of communications) can benefit from the use of diversity.
Although this work largely focuses on Earth-to-satellite channels, our results and findings can be readily translated to all types of FSO fading channels.

We should emphasize that although we study a combined quantum-classical communication system, in what is to follow the vast majority of our analysis concerns the quantum communications aspects. This is because the fragile nature of quantum signals, their inability to be copied, their quantum-only properties, and their complex evolution through a channel lead to a host of complexities absent in classical signalling.

The rest of this paper is as follows. In Section~\ref{Sec:SysModel} we present our system model. In Sections~\ref{EntDist} and~\ref{Sec:QST}  we discuss entanglement distribution and coherent-state transfer, respectively, within our system model. In Section~\ref{Sec:Results} we present our simulation results, and in Section~\ref{Sec:Discussions} we discuss the implications of our results for the real-world diversity-assisted implementation of Gaussian information protocols, such as the standard QKD and SQCC protocols, over FSO fading channels. Section~\ref{Sec:Conclusions} concludes our work.
    
{{{\it Notations:}
Density operators are denoted by $\rho$ (in the main text) or $\hat{\rho}$ (in the Supplementary Material); the quadrature operators are denoted by $\hat{q}$ and $\hat{p}$, respectively.
Other operators  are denoted by uppercase letters without hat, and the adjoint of an operator is denoted by $(\cdot)^\dagger$.
When the Dirac bra–ket notation is used to denote quantum states, a ket (i.e., a column vector) and a bra (i.e., the conjugate transpose of a ket) are denoted by $\ket{\cdot}$ and $\bra{\cdot}$, respectively.
Vectors (that do not denote quantum states) and matrices are denoted in bold face. 
The transpose of a vector or a matrix is denoted by $(\cdot)^{\top}$.
The $2 \times 2$ identity matrix is denoted by $\mathbb{1}_2=\operatorname{diag}(1,1)$, and the Pauli Z matrix is denoted by $\mathbb{Z}_2=\operatorname{diag}(1,-1)$, where $\operatorname{diag}(\mathbf{v})$ denotes a square diagonal matrix with the elements of vector $\mathbf{v}$ on the main diagonal.
The sets of real numbers and complex numbers are denoted by $\mathbb{R}$ and $\mathbb{C}$, respectively. 
The imaginary unit $\sqrt{-1}$ is denoted by $i$, and the absolute value of $z \in \mathbb{C}$ is denoted by $|z|$. The expectation and the variance of a random variable is denoted by $\langle \cdot \rangle$ and $\operatorname{Var}[\cdot]$, respectively.}}

\section{System Model}\label{Sec:SysModel}

\subsection{System Configuration}\label{Sec:CST_General}
\begin{figure}[hbt!]
\centering
\includegraphics[width=0.75\linewidth]{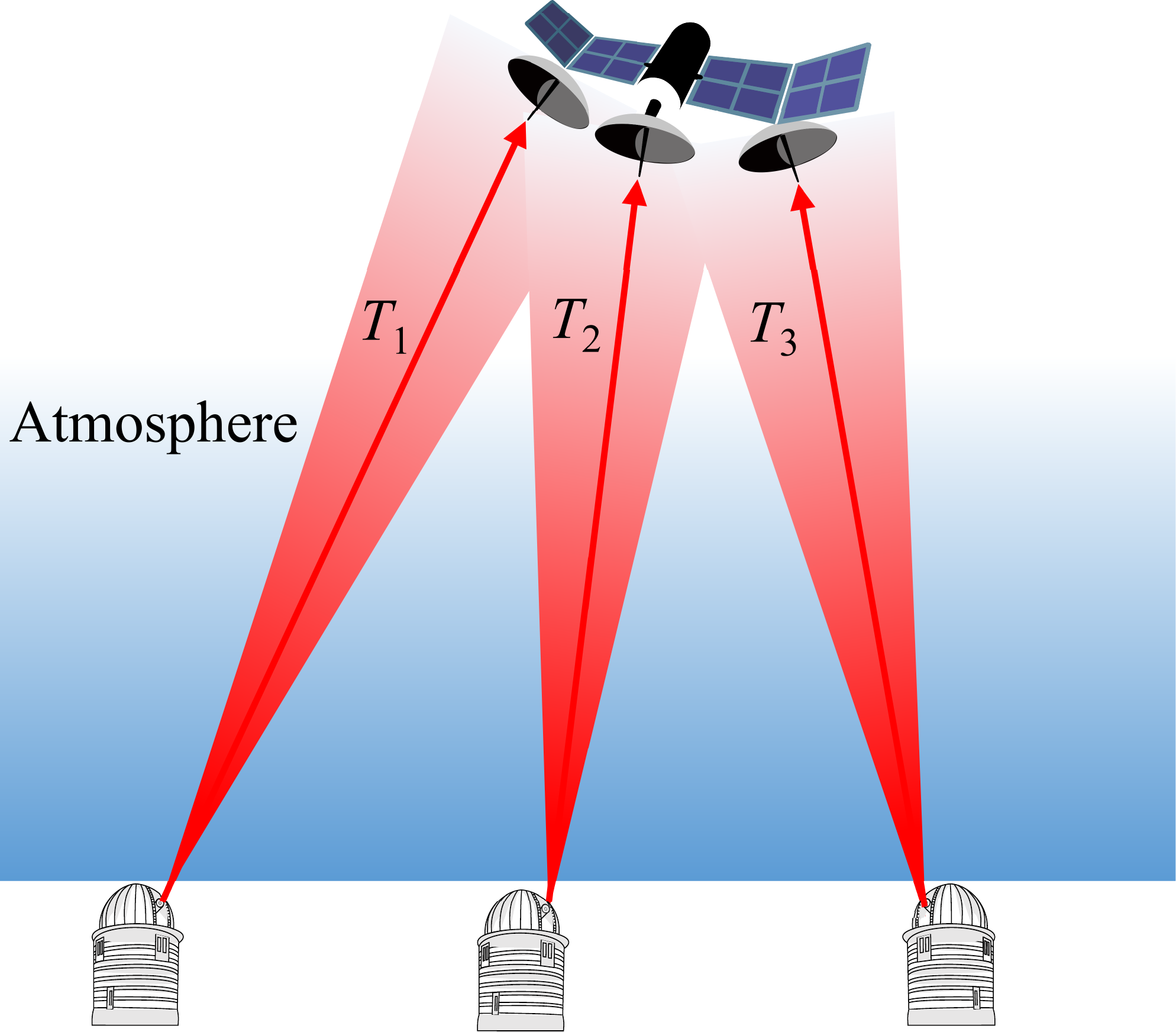}
\caption{System model for our calculations. We assume that only the statistics of the channel is known through some \textit{a priori} channel estimation program.}\label{fig:SysModel}
\end{figure}

As described in Fig.~\ref{fig:SysModel}, in this work we consider an Earth-to-satellite (i.e., uplink) communication system with an $M \times M$ diversity scheme. In this system, $M$ coordinated ground stations (located within proximity on an observatory site) send their corresponding quantum signals to a Low-Earth-Orbit (LEO) satellite employing an array of $M$ receiving apertures onboard. The transmitting ground stations and the receiving satellite apertures are matched in a one-to-one basis -- specifically, transmitting ground station $j$ ($1\le j \le M$) is to send its signal beam to its designated receiving aperture (aperture $j$) at the satellite, and receiving aperture $j$ only expects the signal beam from its corresponding ground station (i.e., ground station $j$). We denote the coordinator of the $M$ ground station as Alice and the receiving satellite as Bob. Unless otherwise stated, we will assume that only the statistics, i.e., the  Probability Density Function (PDF),  of the transmissivity (as a function of channel direction) is known. This greatly simplifies the practical deployment of the communication system, and the required statistics can be achieved through some intensive \textit{a priori} measurement scheme (in our calculations only Bob requires this information).

We define subchannel $j$ as the quantum channel between the transmitter at ground station $j$ and receiving aperture $j$ at the satellite, and we assume perfect tracking between a ground station and its designated receiving aperture at the satellite. 
The use of $M$ transmitting ground stations and $M$ receiving satellite apertures thus results in $M$ physically separate%
\footnote{Although the subchannels can be physically separate, the signal beams from different transmitting ground stations can overlap at the satellite. In this work, we restrict ourselves to the scenario where the impact of such beam overlapping is effectively suppressed via, e.g., careful design and engineering.}
subchannels, and we refer to the number of subchannels $M$ as the diversity number. The transmissivity%
\footnote{Despite adopting the common notation ``loss'', in calculations we mainly use the transmissivity $T$ with the relation $\text{Loss [dB]}=-10\log_{10}T$.}
of subchannel $j$ is denoted as as $T_j$ ($0\le T_j\le1$, $T_j \in \mathbb{R}$), and the excess noise (defined at Bob's side) in subchannel $j$ is denoted as $\epsilon_j=T_j \epsilon_{j}^{\text{A}}$ (where $\epsilon_{j}^{\text{A}}$ denotes the excess noise referring to Alice -- see~\cite{TheoryOfPracticalImplementation}). In this work, we follow~\cite{SebastianRev,NotarnicolaNLA2023} and set $\epsilon_{j}^{\text{A}}$ to a typical value $\epsilon_{j}^{\text{A}}=0.03\,\text{SNU}$ for all subchannels. 

Throughout this work, we assume that all the used subchannels are independent%
\footnote{In order for the subchannels to be independent, the separations amongst the ground-station transmitters have to be much larger than the atmospheric coherence length. 
Since the atmospheric coherence length of the local atmosphere (i.e., several kilometers into the uplink atmospheric channel) at a ground-station is usually very small (our calculations using well-known equations in~\cite{book} give several centimeters), such separations can be easily achieved amongst the ground-station transmitters.} since the transmitted beams from different ground stations evolve independently within their corresponding subchannels. 
As a result, the subchannel transmissivities $\{T_j\}$ can be modeled as independent random variables whose probability distributions are determined by the atmospheric optical turbulence. 
For simplicity, we will assume that all the subchannels are identically distributed, making $\{T_j\}$ independent and identically distributed (i.i.d.) random variables. 
Our channel modelling techniques are detailed in Appendix~\ref{Sec:ChannelModelling} with the fundamental theory of laser propagation within the Earth's turbulent atmosphere presented in Appendix.~\ref{Sec:AtmosProp}. {{It should be noted that the statistical independence between the $M$ subchannels is a result of sufficient separation between the $M$ ground station transmitters. Since the coherence length at the satellite is of the order of $\sim\!1\,\text{km}$, it is practically infeasible to exploit diversity by employing a separation between the $M$ receiving apertures at a single satellite in an uplink configuration~\cite{FSOC_Review_VC}.}}

\subsection{Operational Principles of Diversity-Assisted System}\label{Sec:Diversity}
In this section, we introduce the operational principles of our diversity-assisted scheme for both coherent-state transfer and entanglement distribution. For conciseness, the underlying theoretical background of this work is presented in~\cite{SM}.

In an $M\times M$ diversity-assisted system utilizing $M$ subchannels, instead of directly transmitting a quantum state to Bob over a single channel, Alice equally splits the quantum state to be transmitted into $M$ parts using an array of beam splitters.
She then transfers each of the $M$ parts to one of the $M$ involving ground stations, which, under Alice's coordination, transmits its corresponding part through one of the $M$ subchannels. In this work, we assume the quantum state transfer between Alice and each of the other ground stations is perfect, i.e., achieved with no loss, noise, or decoherence. 
In this sense, ``Alice'' refers to all ground stations.
At the receiving satellite (Bob), the $M$ received signals (received by different receiving apertures) are coherently combined in the optical domain using a combining module consisting of multiple beam splitters. More details regarding the configuration and mathematical description  of such a combining module are provided in Appendix~\ref{Sec:DiversityCombining}.

In a real-world deployment, our diversity-assisted scheme requires accurate time synchronization between Alice and Bob to compensate for the time-of-arrival differences amongst the subchannels signals -- such an accurate synchronization has been demonstrated in the world's first Earth-to-satellite quantum teleportation experiment~\cite{UplinkTeleportation2017}. Our scheme also naturally relies on high-accuracy pointing and tracking due to the high moving speed of an LEO satellite. Failing to address such a requirement will result in a detrimental loss penalty on the system. 
{{We note that the standard deviation of tracking error in the recent proof-of-principle demonstration of uplink teleportation~\cite{UplinkTeleportation2017} was reported to be $<\!3\,\micro\text{rad}$. Our calculations using well-known equations in~\cite{book} suggest that this tracking error is much smaller than the angular displacement of turbulence-induced beam wandering (details not presented here), which is assumed not mitigated in this work.
Furthermore, the accuracy of pointing and tracking is expected to improve with ongoing advancements in the field. Therefore, for simplicity, this work assumes that the impact of pointing and tracking errors is negligible. 
A detailed investigation of pointing and tracking errors would necessitate an accurate modeling of the specific acquisition, pointing, and tracking systems employed at the ground stations (and the satellite). However, we believe such  modeling would not impact our conclusions in any significant manner.}}

Here, we address our assumption of negligible beam-overlapping impact, which, in our scenario of interest, is not only well justified but also represents the worst-case scenario to some extent. 
In a practical implementation, we envision that high-precision tracking (demonstrated in the uplink quantum teleportation experiment~\cite{UplinkTeleportation2017}) is maintained between each pair of transmitting and receiving apertures.
Since the beams from different subchannels have different angles/directions of arrival, from the perspective of each receiving aperture, all the subchannel beams except for one expected beam are tilted in their wavefronts. Since a wavefront tilt is converted into a spot displacement at the focal plane of the receiving optics, the portion of an unexpected beam that illuminates a receiving aperture (due to beam overlapping) cannot be effectively coupled into the system. As a result, beam overlapping at the satellite has a negligible impact on our system.

{{Non-negligible beam-overlapping impact arises when pairwise pointing and tracking cannot be maintained between every transmitting aperture and its corresponding receiving aperture.
For example, if all the ground-station transmitting apertures adopt the same pointing angle (while the receiving apertures at the satellite point in different directions to achieve tracking), the assumption that different subchannels have different angles of arrival is valid only when the ground stations are separated by a large distance that may hinder the effectiveness of quantum-state transfer between Alice and the ground stations. 
The same is true when only one receiving aperture is used at the satellite, and all the transmitting apertures have to point to the same receiving aperture.}}
However, rather than creating undesirable crosstalk noise, beam overlapping effectively couples more portions of signal into the system and allows for a potential performance enhancement since our scheme naturally requires all the subchannel signals to be combined at Bob to reproduce the very signal before Alice's splitting. 
It should be noted that our diversity-assisted scheme, where different transmitters send exactly the same signals derived from an equal splitting of one single source, is fundamentally different from the scheme proposed in~\cite{Sahu2023}, where different transmitters send different signals \textit{independent} of each other.
{{In particular, although the scheme proposed in~\cite{Sahu2023} leverages the full multi-input multi-output processing power (which is not always practically feasible in an uplink configuration), its performance is limited by the negative impact of beam overlapping (if not negligible) and its reliance on knowledge of instantaneous channel transmissivity for proper operation.}}

\section{Entanglement Distribution}\label{EntDist}
For the purpose of distributing entanglement, Alice prepares a sequence of two-mode CV quantum systems in a Two-Mode Squeezed Vacuum (TMSV) state $\rho_{\text{A}{\text{B}}}=\ket{\psi_{\text{AB}}^{\text{TMSV}}}\bra{\psi_{\text{AB}}^{\text{TMSV}}}$ with quadrature variance $V_{\text{s}}$ (which is directly related to its squeezing level) as the entanglement resource (note that A and B denote the two modes). Specifically, Alice keeps mode A at her side and transmits mode B to Bob following the operational principles in Sec.~\ref{Sec:Diversity}. We denote Bob's final output mode (after diversity combining) as mode $\text{B}^{\prime}$, and we denote the final two-mode state shared (i.e., distributed) between Alice and Bob as $\rho_{\text{A}{\text{B}}^{\prime}}$. 
Using this distributed state, Alice and Bob can perform a variety of quantum information protocols, such as Entanglement-Based (EB) CV-QKD. More details on TMSV states can be found in Sec.~\ref{Sec:GaussianStates} of~\cite{SM}.

\subsection{Quantum State Evolution}\label{Sec:QSEvo_TMSV}
We use the statistical-moment formalism (more details can be found in~\cite{SM}) to describe the quantum-state evolution in entanglement distribution. 
In entanglement distribution without using diversity, Alice simply passes mode B of her TMSV state $\rho_{\text{A}{\text{B}}}$ to a ground station. The ground station then transmits mode B through an atmospheric channel (with channel transmissivity $T$ and excess noise $\epsilon$) to the satellite. 
As described in Appendix~\ref{Sec:BS_Channel}, we model the atmospheric channel using a beam splitter with the second input (representing the associated environmental mode) initialized in a thermal state. 
The Covariance Matrix (CM) of the total state (i.e., the TMSV state plus a single-mode thermal state) is given by
\begin{equation}
\begin{aligned}
\mathbf{V}_{\text{tot}}&=\mathbf{V}_{\text{TMSV}} \oplus \underbrace{\mathbf{V}_{\text {thermal}}}_{=:(1+\frac{\epsilon}{1-T}) \mathbb{1}_2}\\
&=\left[\begin{array}{ccc}
V_{\text{s}} \mathbb{1}_2 & \sqrt{V_{\text{s}}^2-1} \mathbb{Z}_2 & 0 \\
\sqrt{V_{\text{s}}^2-1} \mathbb{Z}_2 & V_{\text{s}} \mathbb{1}_2 & 0 \\
0 & 0 & (1+\frac{\epsilon}{1-T}) \mathbb{1}_2
\end{array}\right],
\end{aligned}
\end{equation}
where $V_{\text{s}}$ is the quadrature variance of the TMSV state, $\mathbb{1}_2=\operatorname{diag}(1,1)$, $\mathbb{Z}_2=\operatorname{diag}(1,-1)$, and $\oplus$ denotes a direct sum. After applying the beam splitter transformation $\mathbf{B}(\sqrt{T})$ corresponding to the atmospheric channel and tracing out the associated environmental mode (described in Appendix~\ref{Sec:BS_General}), we can express the CM of the resulting two-mode state $\smash{\rho_{\text{A}{\text{B}}^{\prime}}^{(1)}}$ shared between Alice and Bob as
\begin{equation}\label{Eq:SingleChannelGaussianCM}
\begin{aligned}
\mathbf{V}_{\text{AB}^{\prime}}^{(1)}&=\left(\mathbb{1}_2\oplus\mathbf{B}(\sqrt{T})\right)\mathbf{V}_{\text{TMSV}} \left(\mathbb{1}_2\oplus\mathbf{B}(\sqrt{T})\right)^{\top}\\
&=\left[
\begin{array}{cc}
 V_{\text{s}} \mathbb{1}_2 & \sqrt{T (V_{\text{s}}^2-1)} \mathbb{Z}_2 \\
\sqrt{T (V_{\text{s}}^2-1)} \mathbb{Z}_2 & \left[T (V_{\text{s}}-1)+1 + \epsilon\right] \mathbb{1}_2\\
\end{array}
\right].
\end{aligned}
\end{equation}

Since the {{instantaneous channel transmissivity}} is unknown to Bob, an ensemble average has to be taken on the quantum state $\smash{\rho_{\text{A}{\text{B}}^{\prime}}^{(1)}}$ over the random channel transmissivity $T$. 
As a result, the final (ensemble-averaged) state $\rho_{\text{A}{\text{B}}^{\prime},\text{avg}}^{(1)}$ is a non-Gaussian mixture of Gaussian states obtained from individual channel realizations. 
According to the Wigner-function relationship, the elements of the CM of the ensembles-averaged state $\rho_{\text{A}{\text{B}}^{\prime},\text{avg}}^{(1)}$ is given by the convex sum of the moments of the states $\rho_{\text{A}{\text{B}}^{\prime}}^{(1)}$ under all possible values of channel transmissivity $T$ (see analyses in, e.g.,~\cite{NedaNonGaussian}). In order words, the CM of $\rho_{\text{A}{\text{B}}^{\prime},\text{avg}}^{(1)}$ is given by 
\begin{equation}\label{Eq:SingleChannelNonGaussianCM}
\Scale[0.9]{
\begin{aligned}
\mathbf{V}_{\text{AB}^{\prime},\text{avg}}^{(1)}&=\left[
\begin{array}{cc}
 V_{\text{s}} \mathbb{1}_2 &  \langle \sqrt{T}\rangle \sqrt{V_{\text{s}}^2-1} \mathbb{Z}_2 \\
\langle \sqrt{T}\rangle  \sqrt{V_{\text{s}}^2-1} \mathbb{Z}_2 & \left[
\langle T \rangle (V_{\text{s}}-1)+1 + \langle\epsilon\rangle \right] \mathbb{1}_2\\
\end{array}
\right],
\end{aligned}}
\end{equation}
where $\langle \sqrt{T}\rangle$ denotes the expectation of $\sqrt{T}$.

\begin{widetext}
It can be seen from Eq.~(\ref{Eq:SingleChannelNonGaussianCM}) that a fluctuating channel can be considered as a non-fluctuating channel with an effective transmissivity $\smash{T_{\text{eff}}={\langle\sqrt{T}\rangle}^2}$. Indeed, one can rewrite the CM of $\rho_{\text{A}{\text{B}}^{\prime},\text{avg}}^{(1)}$ in Eq.~(\ref{Eq:SingleChannelNonGaussianCM}) as
\begin{equation}\label{Eq:SingleChannelNonGaussianCMEff}
\mathbf{V}_{\text{AB}^{\prime},\text{avg}}^{(1)}=\left[
\begin{array}{cc}
 V_{\text{s}} \mathbb{1}_2 &  \langle \sqrt{T_{\text{eff}}}\rangle \sqrt{V_{\text{s}}^2-1} \mathbb{Z}_2 \\
\langle \sqrt{T_{\text{eff}}}\rangle  \sqrt{(V_{\text{s}}^2-1)} \mathbb{Z}_2 & \left[
 T_{\text{eff}}(V_{\text{s}}-1) + \operatorname{Var}(\sqrt{T})(V_{\text{s}}-1)+ \langle\epsilon\rangle +1 \right] \mathbb{1}_2\\
\end{array}
\right],
\end{equation}
where ${T_{\text{eff}}}=\langle\sqrt{T}\rangle^2$ is the effective transmissivity and $\operatorname{Var}(X)=\langle X^2 \rangle - {\langle X \rangle}^2$ denotes the variance of $X$. Eq.~(\ref{Eq:SingleChannelNonGaussianCMEff}) reveals that the channel fluctuation results in an extra non-Gaussian noise determined by both the fluctuation-related term $\operatorname{Var}(\sqrt{T})$ and the TMSV quadrature variance $V_{\text{s}}$.
\end{widetext}

The analyses above can be readily extended to  diversity-assisted entanglement distribution using $M\ge 2$ subchannels following the operational principles in Sec.~\ref{Sec:Diversity}. After preparing a TMSV state, equally splitting mode B into $M$ parts, and transferring each part to one of the $M$ ground stations,
Alice shares a $(M+1)$-mode state with the $M$ ground stations. For example, when $M=4$, the CM of this five-mode state is given by
\begin{equation}\label{Eq:M4SplitCM}
\Scale[0.75]{
\mathbf{V}_{\text{split}}^{(4)}=\left[
\begin{array}{ccccc}
V_{\text{s}}\mathbb{1}_2 & \frac{\sqrt{V_{\text{s}}^2-1}}{2} \mathbb{Z}_2 &  \frac{\sqrt{V_{\text{s}}^2-1}}{2} \mathbb{Z}_2 &  \frac{\sqrt{V_{\text{s}}^2-1}}{2} \mathbb{Z}_2 &  \frac{\sqrt{V_{\text{s}}^2-1}}{2} \mathbb{Z}_2  \\
\frac{\sqrt{V_{\text{s}}^2-1}}{2} \mathbb{Z}_2 & \frac{V_{\text{s}}+3}{4} \mathbb{1}_2 & \frac{V_{\text{s}}-1}{4} \mathbb{1}_2 &   \frac{V_{\text{s}}-1}{4} \mathbb{1}_2 &  \frac{V_{\text{s}}-1}{4} \mathbb{1}_2 \\
\frac{\sqrt{V_{\text{s}}^2-1}}{2} \mathbb{Z}_2 &  \frac{V_{\text{s}}-1}{4} \mathbb{1}_2 & \frac{V_{\text{s}}+3}{4} \mathbb{1}_2 &   \frac{V_{\text{s}}-1}{4} \mathbb{1}_2 &  \frac{V_{\text{s}}-1}{4} \mathbb{1}_2 \\
\frac{\sqrt{V_{\text{s}}^2-1}}{2} \mathbb{Z}_2 & \frac{V_{\text{s}}-1}{4} \mathbb{1}_2  & \frac{V_{\text{s}}-1}{4} \mathbb{1}_2 &    \frac{V_{\text{s}}+3}{4} \mathbb{1}_2 &  \frac{V_{\text{s}}-1}{4} \mathbb{1}_2 \\
\frac{\sqrt{V_{\text{s}}^2-1}}{2} \mathbb{Z}_2 & \frac{V_{\text{s}}-1}{4} \mathbb{1}_2 & \frac{V_{\text{s}}-1}{4} \mathbb{1}_2 &   \frac{V_{\text{s}}-1}{4} \mathbb{1}_2 &   \frac{V_{\text{s}}+3}{4} \mathbb{1}_2\\
\end{array}
\right].}
\end{equation}
After each part of mode B is sent though a subchannel to Bob, the resulting total quantum state (with an environmental mode associated with each subchannel taken into account) is a $\smash{(2M+1)}$-mode state. 
By tracing out all the $M$ environmental modes, we can obtain the $\smash{(M+1)}$-mode state after subchannel transmission. 
After taking into account Bob's diversity combining (based on the configuration described in Appendix~\ref{Sec:DiversityCombining}) and tracing out all the unused modes, we can express the CM of the two-mode state $\rho_{\text{A}{\text{B}}^{\prime}}^{(M)}$ shared between Alice and Bob as
\begin{equation}\label{Eq:NonGaussianCombinedCM_Pre}
\mathbf{V}_{\text{AB}^{\prime}}^{(M)}=\left[
\begin{array}{cc}
V_{\text{s}} \mathbb{1}_2 &  C_M \mathbb{Z}_2 \\
C_M  \mathbb{Z}_2 & B_M \mathbb{1}_2\\
\end{array}
\right],
\end{equation}
where
\begin{equation}
\begin{aligned}
C_M &= \frac{\left(\sum_{j=1}^{M}\sqrt{T_j}\right)}{M} \sqrt{V_{\text{s}}^2-1},\,\,\text{and}\\
    B_M &= \frac{\left(\sum_{j=1}^{M}\sqrt{T_j}\right)^2}{M^2} (V_{\text{s}}-1) + \frac{\sum_{j=1}^{M}\epsilon_j}{M} +1.
\end{aligned}
\end{equation}

\begin{widetext}
After further averaging $\rho_{\text{A}{\text{B}}^{\prime}}^{(M)}$ in Eq.~(\ref{Eq:NonGaussianCombinedCM_Pre}) over the joint distribution of the $M$ fluctuating subchannels, we can analytically express the CM of the final (ensemble-averaged) two-mode state $\rho_{\text{A}{\text{B}}^{\prime},\text{avg}}^{(M)}$ shared between Alice and Bob as
\begin{equation}\label{Eq:NonGaussianCombinedCM}
\mathbf{V}_{\text{AB}^{\prime},\text{avg}}^{(M)}=\left[\begin{array}{cc}
V_{\text{s}} \mathbb{1}_2 &\sqrt{T_{\text{eff}}} \sqrt{\left(V_{\text{s}}^2-1\right)} \mathbb{Z}_2 \\
\sqrt{T_{\text{eff}}} \sqrt{\left(V_{\text{s}}^2-1\right)} \mathbb{Z}_2 & {\left[T_{\text{eff}}(V_{\text{s}}-1)+\frac{1}{M} \operatorname{Var}[\sqrt{T}](V_{\text{s}}-1)+\langle \epsilon\rangle +1\right]} \mathbb{1}_2
\end{array}\right],
\end{equation}
where $\smash{{T_{\text{eff}}}=\langle\sqrt{T}\rangle^2}$ is the effective transmissivity same as that in Eq.~(\ref{Eq:SingleChannelNonGaussianCMEff}). Note that, in order to derive Eq.~(\ref{Eq:NonGaussianCombinedCM}), for $\smash{1\le j \le M}$ we have used the relations $\smash{\left\langle T_j\right\rangle=\langle T\rangle}$, $\smash{\left\langle \sqrt{T_j}\right\rangle=\langle \sqrt{T}\rangle}$, $\smash{\operatorname{Var}(\sqrt{T_j})=\operatorname{Var}(\sqrt{T})}$, and $\smash{\langle \epsilon_j\rangle=\langle \epsilon\rangle}$ due to the assumption of i.i.d. subchannels.
\end{widetext}%
For the numerical investigation of diversity-assisted entanglement distribution using our phase-screen simulations of Earth-to-satellite atmospheric channels (to be presented in Sec.~\ref{Sec:EntDistResults}), $\mathbf{V}_{\text{A}{\text{B}}^{\prime},\text{avg}}^{(M)}$ can be evaluated  either by numerically averaging $\mathbf{V}_{\text{A}{\text{B}}^{\prime}}^{(M)}$ in Eq.~(\ref{Eq:NonGaussianCombinedCM_Pre}) over all realizations of the subchannel transmissivities $\{T_j\}$ or by directly plugging in the statistics $\langle{T}\rangle$, $\langle\sqrt{T}\rangle$, and $\operatorname{Var}[\sqrt{T}]$ acquired from our simulations into Eq.~(\ref{Eq:NonGaussianCombinedCM}).

\subsection{Performance Metric}\label{Sec:EntMetrics}
Intuitively, the performance of entanglement distribution is quantified by the  amount of entanglement shared between Alice and Bob after Bob's diversity combining, which is the amount of entanglement contained in the final two-mode state $\smash{\rho_{\text{A}{\text{B}}^{\prime},\text{avg}}^{(M)}}$.
In this work, we adopt two of the most widely used entanglement measures, namely, the \textit{logarithmic negativity} $E_{\text{LN}}$ and the \textit{Reverse Coherent Information (RCI)} $\mathcal{R}_\text{C}$, as our performance metrics for entanglement distribution. 
In particular, we evaluate these two entanglement measures using well-known methods assuming the two-mode state in question is a Gaussian state. 
Although the final two-mode state ${\rho}_{{\text{A}{\text{B}}^{\prime}},\text{avg}}^{(M)}$ shared between Alice and Bob is clearly a non-Gaussian mixed state, such a Gaussian approximation is still justified due to the extremality of Gaussian states -- indeed, it is well-known that for every given CM, the entanglement is minimized by Gaussian states~\cite{ExtremalityOfGaussian}. In other words, our result on the amount of distributed entanglement represents a lower bound on the amount\footnote{Note that,  the efficiently evaluation of the exact amount of entanglement contained in an arbitrary state is a highly non-trivial task.} of entanglement contained in the final two-mode state ${\rho}_{{\text{A}{\text{B}}^{\prime}},\text{avg}}^{(M)}$ shared between Alice and Bob. 
Such a lower bound is commonly referred to as the \textit{Gaussian entanglement}, and it quantifies the amount of entanglement that is useful for Gaussian quantum information protocols such as standard CV-QKD and the teleportation of Gaussian states (see discussions in, e.g.,~\cite{CV_NonGaussianDistillation}).

It can be seen that the CM of the final two-mode state $\rho_{\text{A}{\text{B}}^{\prime},\text{avg}}^{(M)}$ shared between Alice and Bob has the general from
\begin{equation}\label{Eq.TwoModeCMGeneral}
    \mathbf{V}_{{\text{A}{\text{B}}^{\prime},\text{avg}}}^{(M)}=\left[\begin{array}{cc}
a\mathbb{1}_2 & c\mathbb{Z}_2 \\
c\mathbb{Z}_2 & b\mathbb{1}_2
\end{array}\right],
\end{equation}
where $\{a,b,c\}\in\mathbb{R}$. This form greatly facilitates the evaluation of our performance metrics.

\subsubsection{Logarithmic Negativity}
The logarithmic negativity is a widely used entanglement measure that gives an upper bound on the distillable entanglement~\cite{LogNegativity2002,LogNegativity2005}. The logarithmic negativity of a bipartite quantum state $\rho$ is given by 
\begin{equation}
E_{\text{LN}}\left({\rho}\right)=\log_2\left(1+2 \mathcal{N}\left({\rho}\right)\right),
\end{equation}
where $\mathcal{N}({\rho})$ denotes the negativity that  corresponds to the absolute value of the sum of negative eigenvalues of ${\rho}^{\text{PT}}$ (i.e., the partial transpose with respect to either subsystem).
In the spacial cases where ${\rho}$ is a Gaussian state whose CM ${\mathbf{V}}$ follows the general form given by Eq.~(\ref{Eq.TwoModeCMGeneral}), the logarithmic negativity of this state can be simply given by
\begin{equation}\label{Eq:ELN_Gaussian}
    E_{\text{LN}}({\rho})=\max \left[0,-\log _2\left(\nu_{-}\right)\right],
\end{equation}
where $\nu_{-}$ is the smallest symplectic eigenvalue of the partially transposed CM ${\mathbf{V}}$~\cite{GaussianLogNegativity2007}. This symplectic eigenvalue is given by
\begin{equation}
    \nu_{-}=\sqrt{\frac{\Delta-\sqrt{\Delta^2-4 \operatorname{det}({\mathbf{V}})}}{2}},
\end{equation}
where $\Delta=a^2+b^2+2c^2$.
It can be seen that the logarithmic negativity is a function of the quadrature variance $V_{\text{s}}$ of Alice's initial TMSV state. 
In this work, we further scale the evaluated logarithmic negativity of the final state shared between Alice and Bob (i.e., $\smash{E_{\text{LN}}(\rho_{\text{A}{\text{B}}^{\prime},\text{avg}}^{(M)})}$) by the logarithmic negativity of Alice's initial TMSV state $\smash{E_{\text{LN}}({\rho}_{{\text{A}\text{B}}})}$ to better illustrate the evolution of entanglement.
We will refer to this scaled quantity as the scaled logarithmic negativity, and we will use the same symbol $E_{\text{LN}}$ for this scaled quantity when presenting our results.

\subsubsection{RCI}
As another useful entanglement measure that gives a lower bound of the distillable entanglement, the RCI (computed over Alice and Bob's shared state $\rho_{\text{A}{\text{B}}}$) is defined as~\cite{RCI2009,RCICalc2009}
\begin{equation}\label{Eq:RCI_Def}
    \mathcal{R}_{\text{C}} = H\left(\rho_{\text{A}}\right)-H\left(\rho_{\text{A}{\text{B}}}\right),
\end{equation}
where $\smash{\rho_{\text{A}} = \operatorname{tr}_{\text{B}}\left(\rho_{\text{A}{\text{B}}}\right)}$ and $H(\cdot)$ denotes the von Neumann entropy.
In the spacial cases where $\rho_{\text{AB}}$ is a two-mode Gaussian state whose CM follows the general form given by Eq.~(\ref{Eq.TwoModeCMGeneral}), the RCI is simply given by
\begin{equation}
    \mathcal{R}_{\text{C}} = g(\nu^{(\text{A})}) - \sum_{k} g\left(\nu_{k}^{(\text{AB})}\right),
\end{equation}
where $\smash{\nu^{(\text{A})}}$ denotes the symplectic eigenvalue of $\mathbf{V}_{\text{A}}$ (i.e., the CM of $\rho_{\text{A}}$), $\smash{\{\nu_k^{(\text{AB})}\}_{k=1}^{2}}$ denote the symplectic eigenvalues of $\mathbf{V}_{\text{AB}}$ (i.e., the CM of $\rho_{\text{AB}}$), and $\smash{g(x)=\frac{x+1}{2} \log_2 (\frac{x+1}{2}) - \frac{x-1}{2} \log_2 (\frac{x-1}{2})}$. The symplectic eigenvalues $\smash{\{\nu_k\}_{k=1}^{N}}$ of a $2N$-by-$2N$ CM $\mathbf{V}$ can be readily computed as the modulus of the standard eigenvalues of the the matrix $\tilde{\mathbf{V}}=i\boldsymbol{\Omega}\mathbf{V}$, where $\boldsymbol{\Omega}$ is the symplectic form~\cite{GaussianQuantumInformation} given by Eq.~(\ref{Eq:SymplecticForm}) of~\cite{SM}.

The distributed entanglement can be used for various quantum information protocols, the most well-known of which is QKD. 
Generally, in a CV-QKD protocol, after measurement and sifting, Alice and Bob perform parameter estimation, where they assume all the lost signals are acquired by Eve (who controls the channel) and determine the Holevo bound, other relevant parameters, and eventually, the secret key rate from their estimation of the CM in an EB picture~\cite{TheoryOfPracticalImplementation}. In our scenario of interest, none of the estimated CM elements is a function of the instantaneous subchannel transmissivities or noises. Since we do not limit Eve's capabilities in any way, all our analyses and findings are valid under the most powerful attacks allowed by quantum physics.

In this work, we focus on the performance of entanglement distribution rather than the performance of any specific quantum information protocol (that utilizes the distributed entanglement) since entanglement distribution is a  more fundamental application that not only directly supports a variety of quantum information protocols (such as quantum teleportation and different QKD protocols) but also enables the establishment of entanglement between arbitrary network nodes (via entanglement swapping and distillation carried out by quantum repeaters~\cite{QuantumInternetRevModPhys2023}) over a distance much longer than that can be supported by a single satellite, underpinning the emergence of a true global-scale quantum internet.

\section{Coherent-State Transfer}\label{Sec:QST}
Most of the existing classical optical communication systems rely on the transfer of modulated coherent states. {{To a large extent, the same can be said for quantum communication systems. For example, when a TMSV state is used for CV-QKD, the resulting EB CV-QKD protocol is equivalent to a Prepare-and-Measurement (PM) CV-QKD protocol, which relies on the transfer of small quantum-modulated coherent states~\cite{TheoryOfPracticalImplementation}.}}
{{Furthermore, the quantum-classical signaling adopted by the SQCC protocol proposed in~\cite{SQCC2016} relies solely on the modulation of coherent states. As the first step towards the investigation of diversity-assisted SQCC over FSO fading channels we investigate the transfer of large\footnote{{{Note that, in this work we use the term ``small coherent state'' (``large coherent state'') to refer to a coherent state with photon number $\le 10$ ($\ge 100$). We use the term "coherent state" on its own to refer to a coherent state with any photon number.}}}
(classically-modulated) coherent states (for the classical part of SQCC) and small (quantum modulated) coherent states (for the quantum part of SQCC) separately.}}

In general, our diversity-assisted coherent-state transfer aims to transfer the information encoded in the complex amplitude $\alpha_{\text{tar}}\in\mathbb{C}$ (the \textit{encoding amplitude}) of a single-mode coherent state $\ket{\alpha_{\text{tar}}}$ (the \textit{encoded state}) from Alice to Bob following the operational principles in Sec.~\ref{Sec:Diversity}. More details on coherent states can be found in Sec.~\ref{Sec:GaussianStates} of~\cite{SM}.
When large coherent states are used in classical communications, we assume that Alice uses the simplest Binary Phase-shift keying (BPSK) modulation by mapping each of her classical bits according to a \textit{classical alphabet} $\alpha_{\text{tar}}\in\{-\alpha,+\alpha\}$ with $\alpha\in \mathbb{R}$ being a large number. Specifically, $\alpha_{\text{tar}}$ is drawn from a distribution whose PDF is given by
\begin{equation}
    P_{\text{mod}}^{\text{classical}}(\alpha_{\text{tar}})=\begin{cases}
       P_0 \,\,,& \alpha_{\text{tar}}=-\alpha\\
    P_1\,\,,& \alpha_{\text{tar}}=+\alpha
    \end{cases}.
\end{equation}
Without loss of generality, we will assume $P_0=P_1=0.5$ throughout this work.
When small coherent states are used for quantum communications, $\alpha_{\text{tar}}$ is randomly drawn from a continuous \textit{quantum alphabet} each time (see, e.g.,~\cite{RealisticCVTeleportationNG2010}).
To align ourselves with the originally proposed SQCC protocol~\cite{SQCC2016}, in this work we choose the standard PM CV-QKD as our specific application of quantum communication. Specifically, we assume that Alice randomly draws $\alpha_{\text{tar}}$ from a Gaussian distribution whose PDF is given by
\begin{equation}\label{Eq:CSDistribution}
    P_{\text{mod}}^{\text{quantum}}(\alpha_{\text{tar}})=\frac{4}{\pi {V}_{\text{mod}}} \exp \left[-\frac{4|\alpha_{\text{tar}}|^2}{{V}_{\text{mod}}}\right],
\end{equation}
where ${V}_{\text{mod}}$ is the \textit{modulation variance} of the quadrature operators (i.e., $\hat{q}$ and $\hat{p}$) and is commonly referred to as the modulation variance of the standard PM CV-QKD~\cite{TheoryOfPracticalImplementation}. 
{{Note that a standard PM CV-QKD protocol that modulates coherent states with modulation variance $V_{\text{mod}}$ is equivalent to a standard EB CV-QKD protocol using a TMSV state with quadrature variance $V_{\text{s}}=V_{\text{mod}}+1$~\cite{TheoryOfPracticalImplementation}. }}

\subsection{Quantum State Evolution}
We use the Characteristic Function (CF) formalism to describe the quantum state evolution in the process of coherent-state transfer (more details can be found in~\cite{SM}). 
In coherent-state transfer without using diversity, the coherent state $\smash{\ket{\alpha_{\text{tx}}}=\ket{\alpha_{\text{tar}}}}$ is sent through a lossy atmospheric channel with transmissivity $T$ and excess noise $\epsilon$. 
Using the beam-splitter description of the atmospheric channel (see Appendix~\ref{Sec:BS_Channel}), we can express the CF of the output state $\rho^{(1)}$ that reaches the satellite receiver as
\begin{equation}\label{Eq:Fidelity_ST}
\Scale[0.95]{\chi^{(1)}(\xi\,;T,\epsilon,\alpha_{\text{tx}}) = \chi_{\text{coh}}\left(\sqrt{T} \xi\,;\alpha_{\text{tx}}\right) \chi_{\text{vac}}\left(\sqrt{1-T+\epsilon} \,\xi\,;\alpha_{\text{tx}}\right)}.
\end{equation}
When $\smash{M\ge 2}$ subchannels are used in diversity-assisted coherent transfer, the transmitted state in each subchannel becomes $\ket{\alpha_{\text{tx}}^{(M)}}=\ket{\alpha_{\text{tar}}/\sqrt{M}}$ due to Alice's equal splitting (recall Sec.~\ref{Sec:Diversity}). After the  transmission over subchannels and Bob's diversity combining based on the configuration described in Appendix~\ref{Sec:DiversityCombining}, the CF of Bob's final state can be readily expressed using the beam-splitter description detailed in Appendix~\ref{Sec:BS}.

For illustration purposes, in the following we explicitly present the CF of Bob's final combined state in $\smash{M \times M}$ diversity-assisted coherent-state transfer with $M$ ranging from 2 to 4.
When $M=2$, the combining module consists of only one beam splitter (see Fig.~\ref{Fig.Combining}(a) in Appendix~\ref{Sec:DiversityCombining}), and the CF of the final state $\rho_{\text{DC}}^{(2)}$ is given by
\begin{equation}\label{Eq:Fidelity_D2}
    \Scale[0.8]{\chi_{\text{DC}}^{(2)}(\xi) = \chi^{(1)}\left(\eta_1\xi\,;T_1,\epsilon_1,\alpha_{\text{tx}}^{(2)}\right) 
    \chi^{(1)}\left(\sqrt{1-\eta_1^2}\xi\,;T_2,\epsilon_2,\alpha_{\text{tx}}^{(2)}\right)}.
\end{equation}
When $M=3$, the combining module consists of two beam splitters (see Fig.~\ref{Fig.Combining}(b) in Appendix~\ref{Sec:DiversityCombining}), and the CF of the final state $\rho_{\text{DC}}^{(3)}$ is given by
\begin{equation}\label{Eq:Fidelity_D3}
\Scale[0.8]{\begin{aligned}
    \chi_{\text{DC}}^{(3)}(\xi)=\,&\chi^{(1)}\left(\eta_1\eta_2\xi\,;T_1,\epsilon_1,\alpha_{\text{tx}}^{(3)}\right) 
    \chi^{(1)}\left(\eta_2\sqrt{1-\eta_1^2}\xi\,;T_2,\epsilon_2,\alpha_{\text{tx}}^{(3)}\right)\\
    &\chi^{(1)}\left(\sqrt{1-\eta_2^2}\xi\,;T_3,\epsilon_3,\alpha_{\text{tx}}^{(3)}\right).
\end{aligned}}
\end{equation}
When $M=4$, the combining module consists of three beam splitters (see Fig.~\ref{Fig.Combining}(c) in Appendix~\ref{Sec:DiversityCombining}), and the CF of the final state $\rho_{\text{DC}}^{(4)}$ is given by
\begin{equation}\label{Eq:Fidelity_D4}
\Scale[0.8]{\begin{aligned}
\chi_{\text{DC}}^{(4)}(\xi)=\,&\chi^{(1)}\left(\eta_1\eta_3\xi\,;T_1,\epsilon_1,\alpha_{\text{tx}}^{(4)}\right) 
\chi^{(1)}\left(\eta_3\sqrt{1-\eta_1^2}\xi\,;T_2,\epsilon_2,\alpha_{\text{tx}}^{(4)}\right)\\
&\chi^{(1)}\left(\eta_2\sqrt{1-\eta_3^2}\xi\,;T_3,\epsilon_3,\alpha_{\text{tx}}^{(4)}\right)\\
&\chi^{(1)}\left(\sqrt{\left(1-\eta_2^2\right)\left(1-\eta_3^2\right)}\xi\,;T_4,\epsilon_4,\alpha_{\text{tx}}^{(4)}\right).
\end{aligned}}
\end{equation}

\subsection{Performance Metric}\label{Sec:MetricCT}
The \textit{fidelity} quantifies the closeness between two quantum states.
Under the CF formalism, the definition of fidelity between two quantum states (denoted as $\rho_1$ and $\rho_2$) is given by
\begin{equation}\label{Eq:FidelityCF}
    \mathcal{F}(\rho_1,\rho_2)=\frac{1}{\pi} \int_{\mathbb{R}^2}  \chi_{1}(\xi) \chi_{2}(-\xi)\,\mathrm{d}^2\xi,
\end{equation}
where $\chi_{1}(\xi)$ ($\chi_{2}(\xi)$) denotes the CF of $\rho_1$ ($\rho_2$)~\cite{RealisticCVTeleportationNG2010}. 
Naturally, the effectiveness of transferring any given encoded state $\ket{\alpha_{\text{tar}}}$ from Alice to Bob can be quantified by the fidelity between Bob's final state and a scaled version of Alice's encoded state $\smash{\ket{\alpha_{\text{tar}}^{\prime}}=\left|{\alpha_{\text{tar}}/{\langle \sqrt{T}\rangle}}\right\rangle}$. 
Specifically, the scaling factor $\smash{{1}/{\langle \sqrt{T}\rangle}}$ is applied to reflect the necessary operation in a real-world system -- Bob has to scale his measurement results (using the available information of channel statistics) to compensate for the (fixed) mean channel loss.  
Under one realization of the $M$ fluctuating subchannel transmissivities, such a fidelity can be calculated as $\smash{\mathcal{F}\left(\rho_{\text{DC}}^{(M)}, \ket{\alpha_{\text{tar}}^{\prime}}\bra{\alpha_{\text{tar}}^{\prime}}\right)}$ using Eq.~(\ref{Eq:FidelityCF}). The analytical expression of this fidelity is presented in Appendix~\ref{A:AnalyticalFidelity}.

In this work, we use the \textit{average fidelity} as the performance metric for coherent-state transfer. Note that the term \textit{average} here refers to taking average over both the fluctuating channel transmissivities and the (classical or quantum) alphabet.
Since Bob does not possess the knowledge of {{instantaneous channel transmissivities}}, his final state is a (non-Gaussian) mixed state resulting from an ensemble average over the fluctuating subchannel transmissivities $\{T_j\}$. 
The definition of fidelity allows us to first calculate the non-averaged fidelity $\mathcal{F}$ using Eq.~(\ref{Eq:FidelityCF}) and then take average over the (classical or quantum) alphabet and the fluctuating subchannel transmissivities $\{T_j\}$.
The average fidelity can be explicitly given by
\begin{equation}\label{Eq.AvgFidelity}
\begin{aligned}
\mathcal{F}_{\text{avg}}^{(M)}=&\int_{\mathbb{R}^{M}}  \int_{\mathbb{R}^2}  P_{\text{chn}}(\mathbf{T}) P_{\text{mod}}^{\text{classical/quantum}}(\alpha_{\text{tar}})\\
&\mathcal{F}\left(\rho_{\text{DC}}^{(M)}, \ket{\alpha_{\text{tar}}^{\prime}}\bra{\alpha_{\text{tar}}^{\prime}}\right)\,{{\rm{d}}^2}\,\alpha_{\text{tar}}\,\,{{\rm{d}}^{M}}\,\mathbf{T},
\end{aligned}
\end{equation}
where $P_{\text{chn}}(\mathbf{T})$ denotes the joint PDF of subchannel transmissivities $\{T_j\}$ (defined in Appendix.~\ref{Sec:AtmosProp}).

\section{Simulation Results}\label{Sec:Results}
In this section, we present our main simulation results.
As discussed in Sec.~\ref{Sec:CST_General}, the fluctuating subchannel transmissivities $\{T_j\}$ are i.i.d. random variables whose probability distributions are determined by the optical turbulence within the Earth’s turbulent atmosphere. 
To provide an intuitive measure, despite using transmissivity for all our calculations, for illustration purposes we choose to describe the statistics of a fading channel in the dB domain -- specifically, we quantify the \textit{mean channel loss} using the average value of the fluctuating loss (in dB) and the \textit{channel fading strength} using the standard deviation of the fluctuating loss (also in dB) in any given subchannel.

\subsection{Entanglement Distribution}\label{Sec:EntDistResults}
We now examine the effectiveness of incorporating diversity for the purpose of entanglement distribution. 
{{Specifically, we will plot the performance of diversity-assisted entanglement distribution achieved using initial TMSV states with different values of the quadrature variance $V_{\text{s}}$. 

When investigating the effectiveness of diversity-assisted entanglement distribution, we incorporate our detailed  phase-screen simulations (based on the split-step method) in order to faithfully capture the effects imposed by the optical turbulence on an optical beam propagating within realistic Earth-to-satellite channels (more details are provided in Appendix~\ref{Sec:PhsDetails}).
Our simulation settings are detailed in Appendix~\ref{Sec:SimulationSettings}.
The simulated PDFs and statistics of the fluctuating loss over the typical Earth-to-satellite channels considered are presented in Fig.~\ref{fig:PDF_Loss} and Table~\ref{Table:LossStats} in Appendix~\ref{Sec:PhsDetails}.
}
} 

In Fig.~\ref{fig:ResultsENT_Phs} we adopt all the settings in Appendix~\ref{Sec:SimulationSettings} and plot the scaled logarithmic negativity $E_{\text{LN}}^{(M)}$ achieved by diversity-assisted entanglement distribution under different zenith angles (${\theta_{\text{z}}=0^{\circ}}$, ${\theta_{\text{z}}=30^{\circ}}$, and ${\theta_{\text{z}}=45^{\circ}}$)  in a real-world uplink quantum communication system. 
From Fig.~\ref{fig:ResultsENT_Phs}, we first observe that the TMSV state with a larger quadrature variance $V_{\text{s}}$ (which equivalently indicates a stronger squeezing level) is more susceptible to fading -- such an observation is consistent with the findings in previous works, (e.g.,~\cite{NedaEntDistSat2015}).
It can also be clearly seen that the incorporation of diversity improves the effectiveness of entanglement distribution, and the use of more subchannels leads to performance improvement in terms of logarithm negativity. 
However, the performance improvement provided by adding subchannels is reduced as the number of subchannels increases. This ``diminishing return'' can be explained by Eq.~(\ref{Eq:NonGaussianCombinedCM}) -- indeed, when $M$ increases above a certain threshold, the non-Gaussian noise term $\frac{1}{M}\operatorname{Var}[\sqrt{T}](V_{\text{s}}-1)$ becomes negligible, making the total noise dominated by the term $(\langle \epsilon \rangle+1-T_{\text{eff}})$, which cannot be reduced by the use of diversity. 
Furthermore, the results in Fig.~\ref{fig:ResultsENT_Phs} also indicate that using diversity enables entanglement distribution over subchannels that are entanglement-breaking by themselves. For example, from Fig.~\ref{fig:ResultsENT_Phs} we can see that when $V_{\text{s}}=9$, it is impossible to distribute entanglement without using diversity under zenith angle $\smash{\theta_{\text{z}}=30^{\circ}}$ (see the $\smash{M=1}$ data point of the black dashed curve); however, the use of $2\times 2$ diversity enables the distribution of non-zero entanglement under the same channel condition (see the $\smash{M=2}$ data point of the black dashed curve).
\begin{figure}[htbp!]
\centering
\includegraphics[width=0.95\linewidth]{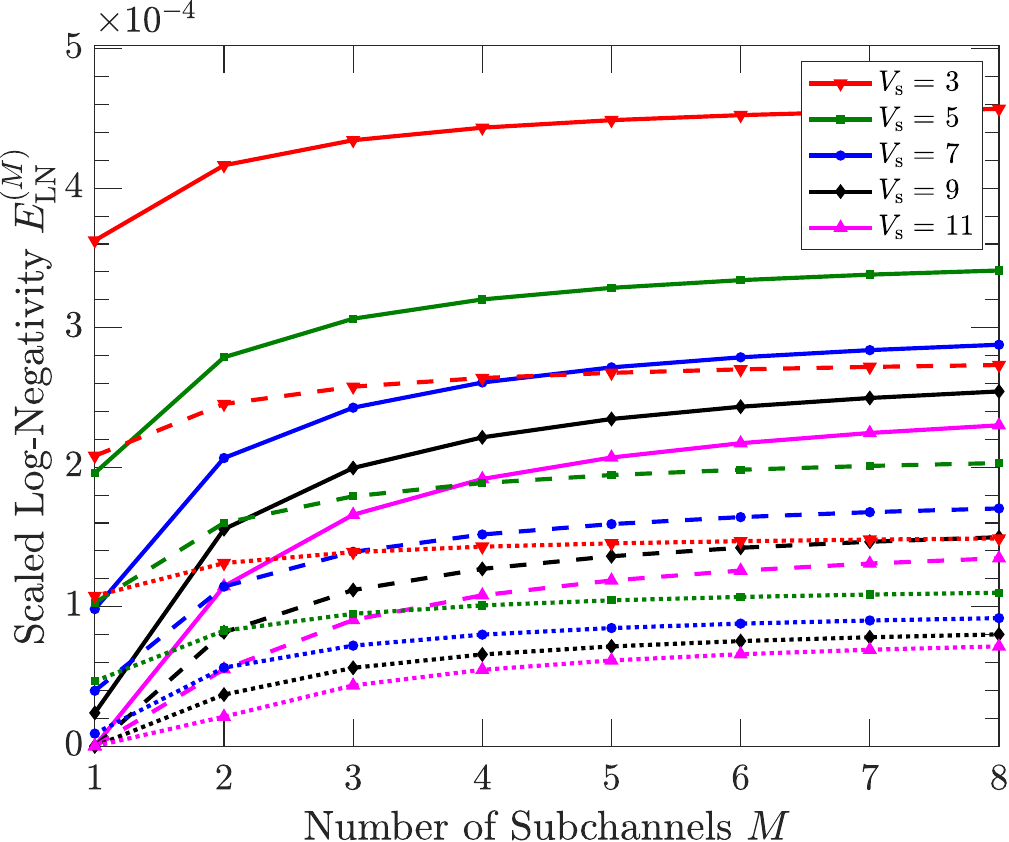}
\caption{{{Scaled logarithmic negativity $ E_{\text{LN}}^{(M)}$ achieved by diversity-assisted entanglement distribution under different zenith angles ${\theta_{\text{z}}=0^{\circ}}$ (solid), ${\theta_{\text{z}}=30^{\circ}}$ (dashed), and ${\theta_{\text{z}}=45^{\circ}}$ (dotted) in a real-world uplink quantum communication system. The results are plotted against the number of subchannels $M$, and different values of the quadrature variance $V_{\text{s}}$ of Alice's initial TMSV state are considered.
All subchannels are modelled via our phase-screen simulations, with the excess noise referring to Alice set to ${\epsilon_j^{\mathrm{A}}=0.03\,\text{SNU}}$. The statistics of the subchannel loss in terms of $\{\text{Mean Loss},\,\text{Fading Strength}\}$ are 
$\{35.2\,\text{dB},\,5.8\,\text{dB}\}$ under ${\theta_{\text{z}}=0^{\circ}}$,
$\{37.6\,\text{dB},\,6.2\,\text{dB}\}$ under ${\theta_{\text{z}}=30^{\circ}}$, and
$\{40.4\,\text{dB},\,6.4\,\text{dB}\}$ under ${\theta_{\text{z}}=45^{\circ}}$ (the corresponding PDFs of loss are presented in Fig.~\ref{fig:PDF_Loss}). Note that the $M=1$ data points represent the performance achieved without using diversity and are plotted as a benchmark.}}}\label{fig:ResultsENT_Phs}
\end{figure}

{{The RCI not only gives a lower bound for the distillable entanglement but also represents a lower bound for the achievable secret key rate when the distributed entanglement is used for QKD. In CV-QKD, the modulation variance $V_{\text{mod}}$ (in a PM implementation), or, equivalently, the corresponding TMSV-state quadrature variance $V_{\text{s}}$ (in an EB implementation), is typically considered a free parameter chosen by Alice and Bob. 
Although the RCI is an information-theoretic metric that does not necessarily indicate the performance of a specific quantum information protocol (recall that the distributed entanglement can be used for purposes other than QKD), given the relevance of the RCI to QKD, one would intuitively be interested in optimizing $V_{\text{s}}$ for a maximized RCI. 
In Fig.~\ref{Fig:RCI_Vs}, we adopt all the settings in Appendix~\ref{Sec:SimulationSettings} and plot the RCI $\mathcal{R}_{\text{C}}^{(M)}$ achieved by diversity-assisted entanglement distribution under different zenith angles (${\theta_{\text{z}}=0^{\circ}}$, ${\theta_{\text{z}}=30^{\circ}}$, and ${\theta_{\text{z}}=45^{\circ}}$) in a real-world uplink quantum communication system. 
Although the main focus of this work is not the performance optimization for any specific quantum information protocol (that utilizes the distributed entanglement), instead of plotting the RCI against the discrete number of subchannels $M$ (as we did for the logarithmic negativity in Fig.~\ref{fig:ResultsENT_Phs}), in Fig.~\ref{Fig:RCI_Vs} we plot the RCI against the TMSV-state quadrature variance $V_{\text{s}}$, providing more useful insights by revealing the relationship between $V_{\text{s}}$ and the RCI in diversity-assisted entanglement distribution. 
\begin{figure}[htbp!]
\centering
\includegraphics[width=0.95\linewidth]{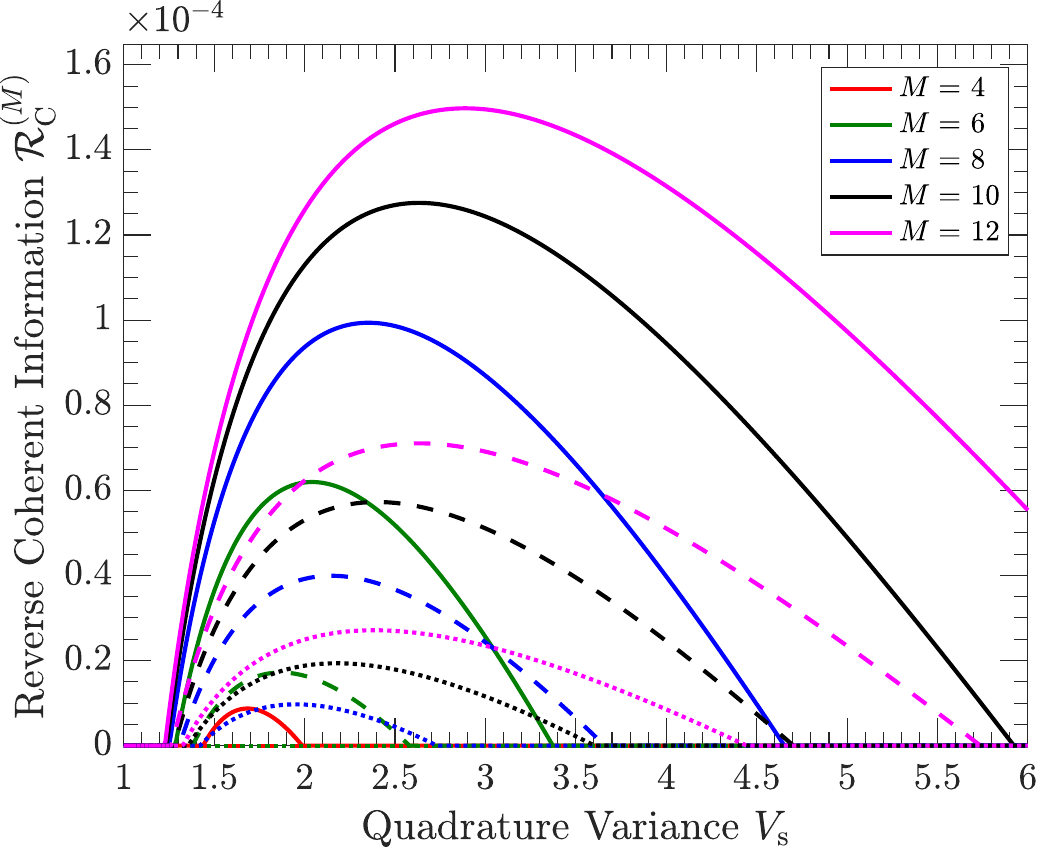}
\caption{{{RCI $\mathcal{R}_{\text{C}}^{(M)}$ achieved by diversity-assisted entanglement distribution under different zenith angles ${\theta_{\text{z}}=0^{\circ}}$ (solid), ${\theta_{\text{z}}=30^{\circ}}$ (dashed), and ${\theta_{\text{z}}=45^{\circ}}$ (dotted) in a real-world uplink quantum communication system. The results are plotted against the quadrature variance of Alice's initial TMSV state $V_{\text{s}}$, and different numbers of subchannels $M$ are considered.
All subchannels are modelled via our phase-screen simulations, with the excess noise referring to Alice set to ${\epsilon_j^{\mathrm{A}}=0.03\,\text{SNU}}$. The statistics of the subchannel loss in terms of $\{\text{Mean Loss},\,\text{Fading Strength}\}$ are 
$\{35.2\,\text{dB},\,5.8\,\text{dB}\}$ under ${\theta_{\text{z}}=0^{\circ}}$,
$\{37.6\,\text{dB},\,6.2\,\text{dB}\}$ under ${\theta_{\text{z}}=30^{\circ}}$, and
$\{40.4\,\text{dB},\,6.4\,\text{dB}\}$ under ${\theta_{\text{z}}=45^{\circ}}$ (the corresponding PDFs of loss are presented in Fig.~\ref{fig:PDF_Loss}).}}}\label{Fig:RCI_Vs}
\end{figure}

From Fig.~\ref{Fig:RCI_Vs}, we first observe that most of the findings from Fig.~\ref{fig:ResultsENT_Phs} (related to the logarithmic negativity) also apply to the RCI, although some may be less apparent since the RCI is plotted against a different variable.
Most importantly, it can be clearly seen that the use of diversity can improve the effectiveness of entanglement distribution in terms of the RCI -- indeed, for a given $V_{\text{s}}$, the use of more subchannels (i.e., a larger $M$) can lead to a higher RCI under a given zenith angle $\theta_{\text{z}}$. On the contrary, without using diversity (i.e., when $M=1$), positive RCI cannot be achieved under any considered channel condition, regardless of the quadrature variance of the initial TMSV state $V_{\text{s}}$ (not shown). 

Fig.~\ref{Fig:RCI_Vs} also reveals that under a given zenith angle $\theta_{\text{z}}$, there exists an optimal quadrature variance 
$V_{\text{s}}$ that gives the maximum RCI within the viable range of  $V_{\text{s}}$ (in which a positive RCI can be achieved) for a given number of subchannels $M$. 
This can be understood with reference to the definition of the RCI in Eq.~(\ref{Eq:RCI_Def}). Specifically, increasing the quadrature variance $V_{\text{s}}$ increases both $H(\rho_{\text{A}})$ and $H(\rho_{\text{AB}})$, but in different ways, resulting in the concave shape observed in the plotted curves. 
In general, the viable range of $V_{\text{s}}$ becomes narrower, and the maximum RCI becomes smaller under a larger $\theta_{\text{z}}$ due to the higher mean loss and stronger fading encountered, but the use of more subchannels broadens the viable range of $V_{\text{s}}$ and allows for a higher maximum RCI to be achieved. 
For example, under ${\theta_{\text{z}}=30^\circ}$, no positive RCI can be achieved when ${M=4}$ subchannels are used (indicated by the absence of the red dashed curve); however, positive RCI can be achieved when ${M=6}$ subchannels are used under the same channel condition (indicated by the green dashed curve). 
Similar to the diversity-enabled improvement in the logarithmic negativity, the diversity-enabled improvement in the RCI is also attributed to the $M$-time reduction of the fluctuation-induced non-Gaussian noise $\frac{1}{M} \operatorname{Var}[\sqrt{T}]\left(V_{\mathrm{s}}-1\right)$.}}

Here, we highlight the difference between our $\smash{M \times M}$ diversity-assisted entanglement distribution and the multiplexing of entanglement distribution over $M$ subchannels. 
When used alone, an entanglement-breaking channel cannot support effective entanglement distribution since Alice and Bob will end up sharing a separable state that does not contain any entanglement. 
Suppose the entanglement distribution is simply multiplexed over $M$ independent subchannels that are statistically identical to an entanglement-breaking channel. In that case, the total amount of distributed entanglement will still be zero since Alice and Bob will end up sharing $M$ separable states. 
As a result of the $M$-time reduction of the fluctuation-induced non-Gaussian noise, the unique performance enhancement that allows for entanglement distribution over entanglement-breaking subchannels is enabled by the use of $M \times M$ diversity and cannot be achieved by any spatial multiplexing scheme known to date.

\subsection{Coherent-State Transfer}\label{Sec:ResultsCST}
We now examine the effectiveness of incorporating diversity for the purpose of coherent-state transfer within the context of SQCC. 
In addition to fading, the high mean loss over Earth-to-satellite channels can be very challenging for the faithful transfer of coherent states  -- indeed, our simulations indicate that the mean loss over a real-world Earth-to-satellite channel can be as high as ${\sim 40\,\text{dB}}$ (see Fig.~\ref{fig:PDF_Loss} and Table~\ref{Table:LossStats} in Appendix~\ref{Sec:PhsDetails} for details).
In our investigation, we find that such a high mean loss essentially renders the faithful transfer of coherent states non-viable.
It should be noted that this observation does not necessarily negate the possibility of classical communications, which focus on discriminating one coherent state from the other rather than the exact transfer of coherent states. Nevertheless, as in the case of coherent-state transfer, such a high loss will likely restrict the viability of the diversity-assisted discrimination of coherent states over fading channels (which has been investigated in~\cite{TCommsDiversityQuantumComms2020}) as well.
In order to demonstrate the usefulness of diversity as a means of fading mitigation in coherent-state transfer over other generic fading channels, we restrict ourselves to a regime with low mean subchannel losses and adopt a simple and flexible log-normal distribution (detailed in Appendix~\ref{Sec:LNDetails}) to investigate the general behaviours of our diversity-assisted system over fading channels.
Our focus on such a low-mean-loss regime makes our results (on average fidelity) more likely to be relevant to other FSO fading channels, such as those considered in~\cite{TCommsDiversityQuantumComms2020} or satellite-based downlink channels that have a low mean loss and insignificant fading.

\begin{figure}[htbp!]
\centering
\includegraphics[width=0.95\linewidth]{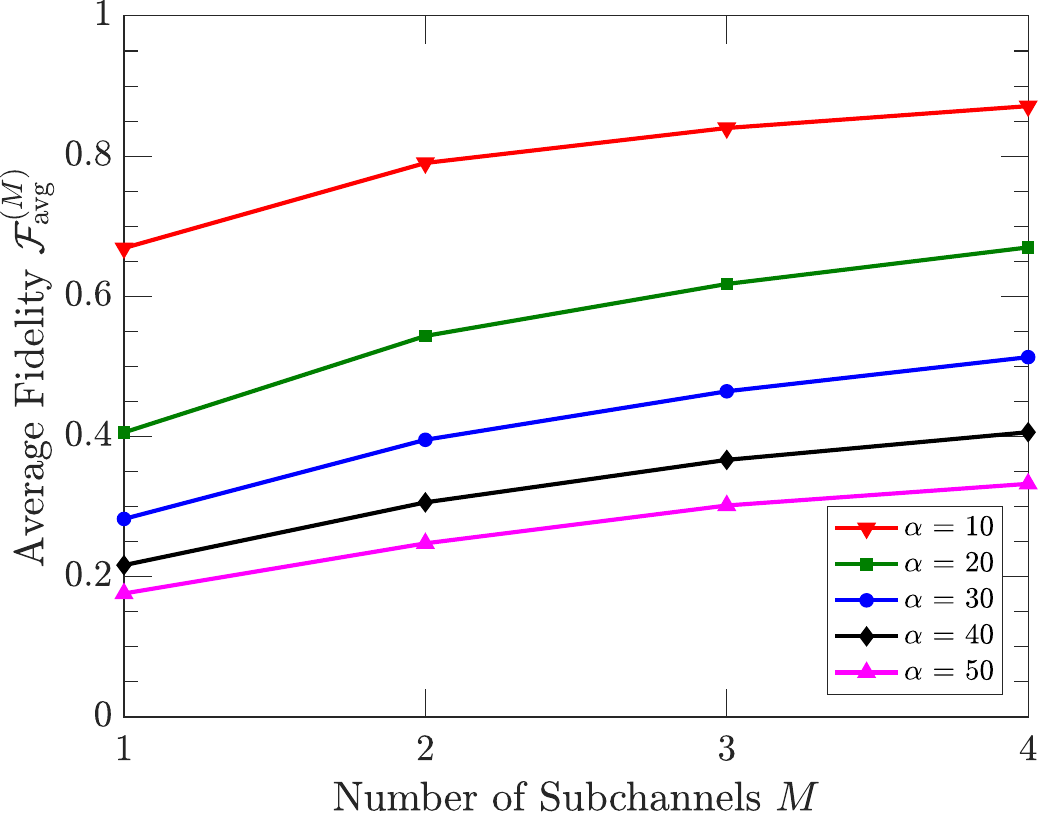}
\caption{{{Performance (average fidelity) of the diversity-assisted transfer of BPSK-modulated large coherent states with $\alpha$ ranging from $10$ ($100$ photons) to $50$ ($2500$ photons). All the subchannels are modelled using our log-normal model, with the excess noise referring to Alice set to $\epsilon_j^{\mathrm{A}}=0.03\,\text{SNU}$. The mean subchannel loss and the fading strength are set to $3\,\text{dB}$ and $1\,\text{dB}$, respectively. Note that the $M=1$ data points represent the performance achieved without using diversity and are plotted as a benchmark.}}} \label{fig:ResultsFixedCS_LN31}
\end{figure}

In Fig.~\ref{fig:ResultsFixedCS_LN31}, we plot the performance (i.e., average fidelity) of the diversity-assisted transfer of BPSK-modulated large coherent states with $\alpha$ ranging from $10$ ($100$ photons) to $50$ ($2500$ photons) achieved under 3 dB mean subchannel loss and 1 dB subchannel fading strength. From this figure, we can clearly see that the use of diversity improves the transfer of large coherent states -- such an observation is consistent with the findings in~\cite{TCommsDiversityQuantumComms2020}, which suggest the usefulness of incorporating diversity within the context of discriminating BPSK-modulated coherent states over fading channels. {{Interestingly, from Fig.~\ref{fig:ResultsFixedCS_LN31} we can also see that the average fidelity of transferring BPSK-modulated coherent states $\smash{\left\{\ket{-\alpha},\ket{+\alpha}\right\}}$ decreases as the amplitude $\alpha$ increases, which indicates that a larger coherent state is more susceptible to channel fading. Such an observation is somewhat in contrast to one in~\cite{TCommsDiversityQuantumComms2020} (i.e., a larger $\alpha$ improves the discrimination of two BPSK-modulated coherent states), and this is because our performance metric for coherent-state transfer (i.e., the average fidelity) focuses on the exact transfer of coherent states rather than discriminating one coherent state from the other.}}

\begin{figure}[htbp!]
\centering
\includegraphics[width=0.95\linewidth]{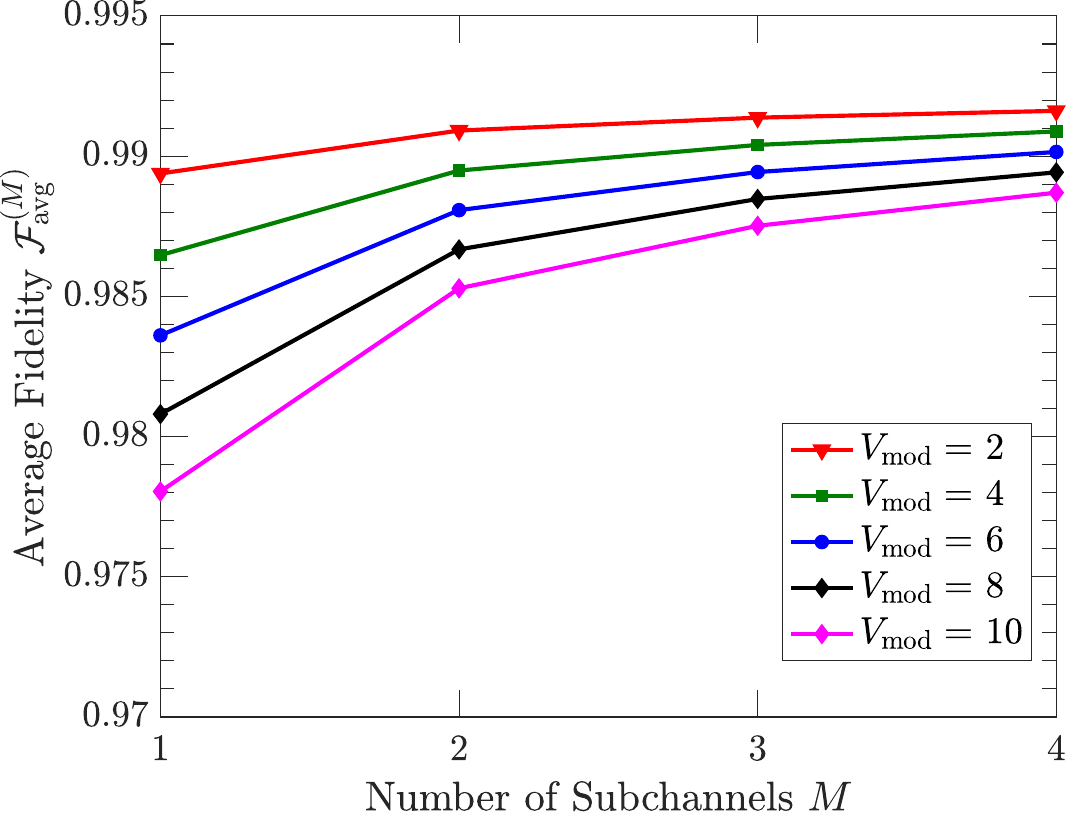}
\caption{{{Performance (average fidelity) of the diversity-assisted transfer of small quantum-modulated coherent states with typical values of modulation variance $V_{\text{mod}}$. All the subchannels are modelled using our log-normal model, with the excess noise referring to Alice set to $\epsilon_j^{\mathrm{A}}=0.03\,\text{SNU}$. The mean subchannel loss and the fading strength are set to $3\,\text{dB}$ and $1\,\text{dB}$, respectively. Note that the $M=1$ data points represent the performance achieved without using diversity and are plotted as a benchmark.}}}\label{fig:ResultsQMCS_LN31}
\end{figure}

In Fig.~\ref{fig:ResultsQMCS_LN31}, we plot the performance (i.e., average fidelity) of the diversity-assisted transfer of small quantum-modulated coherent states with typical values of modulation variance $V_{\text{mod}}$ achieved under 3 dB mean subchannel loss and 1 dB subchannel fading strength. 
{{Since a standard PM CV-QKD protocol (which relies on the transfer of small quantum-modulated coherent states with modulation variance $\smash{V_{\text{mod}}}$) is equivalent to an EB CV-QKD protocol (which relies on the entanglement distribution utilizing a TMSV state with quadrature variance $V_{\text{s}}=V_{\text{mod}}+1$), from the perspective of CV-QKD we set $\smash{V_{\text{mod}}=\{2,4,6,8,10\}}$ to maintain maximum relevance between the transfer of small quantum-modulated coherent states (investigated here) and the distribution of entanglement (investigated in Sec.~\ref{Sec:EntDistResults}).}}
{{From Fig.~\ref{fig:ResultsQMCS_LN31}, we can observe that the average fidelity of transferring small quantum-modulated coherent states decreases as the modulation variance $V_{\text{mod}}$ increases, and this is again because a larger coherent state is more susceptible to channel fading.}}
Most importantly, Fig.~\ref{fig:ResultsQMCS_LN31} clearly indicates that the use of diversity can improve the transfer of small quantum-modulated coherent states over fading channels. 

\section{Discussions}\label{Sec:Discussions}
We have confirmed the usefulness of diversity for both entanglement distribution and coherent-state transfer over fading channels. 
Most straightforwardly, our findings indicate that employing diversity as a fading mitigation method can enhance the performance of all Gaussian quantum information protocols (including the standard CV-QKD) over fading channels (including Earth-to-satellite channels).
{{As mentioned in Sec.~\ref{Sec:CST_General}, in this work we have assumed that only the statistics of the transmissivity is known -- in other words, the instantaneous channel transmissivity, sometimes referred to as the full Channel State Information (CSI), is not available to Bob.
Interestingly, when the full CSI is available at Bob, we find that the diversity-assist system described in Sec.~\ref{Sec:Diversity} does not provide any non-negligible improvement in both entanglement distribution and coherent-state transfer under all our considered scenarios (results not presented).}}

The negligible improvement in entanglement distribution (when the full CSI is available at Bob) can be explained by extending our analyses in Sec.~\ref{Sec:QSEvo_TMSV}. Specifically, the availability of the full CSI at Bob eliminates the non-Gaussian mixture and its resulting fluctuation-related non-Gaussian noise term $\smash{\operatorname{Var}[\sqrt{T}](V_{\text{s}}-1)}$ in Eq.~(\ref{Eq:SingleChannelNonGaussianCMEff}) (when diversity is not used) and Eq.~(\ref{Eq:NonGaussianCombinedCM}) (when diversity is used). 
Since the enhancement provided by using diversity solely comes from the reduction of the aforementioned non-Gaussian noise term (recall the $1/M$ factor in Eq.~(\ref{Eq:NonGaussianCombinedCM})), it is intuitive that the use of diversity cannot perceptibly improve the lower bound of the distributed entanglement when Bob has the full CSI. 
Because such a lower bound (i.e., the Gaussian entanglement) quantifies the amount of entanglement useful for Gaussian quantum information protocols, it should be expected that the real-world implementation of a Gaussian quantum information protocol over fading channels will be unlikely to benefit from the use of diversity in the presence of the full CSI.

The reason for the negligible improvement in coherent-state transfer (when the full CSI is available at Bob) is rather intuitive. Specifically, 
in an implementation that does not employ diversity, Bob can always scale his measurement results using the instantaneous channel transmissivity. As a result, the faithful transfer of classical or quantum information (encoded in coherent states) between Alice and Bob is only limited by the factors (e.g., the excess noise and the sensitivity of Bob's detector) that remain unchanged regardless of whether diversity is employed.

This work is originally motivated by~\cite{TCommsDiversityQuantumComms2020}, which reveals the usefulness of diversity in the discrimination of two BPSK-modulated coherent states over generic log-normal fading channels and thus suggests that the classical communication part of SQCC can benefit from using diversity. {{Indeed, our results on the transfer of classically-modulated coherent states over fading channels modeled in a similar fashion (recall Sec.~\ref{Sec:ResultsCST}) have confirmed this finding from a different perspective.}}

We now discuss the implications of our findings for the practical implementation of diversity-assisted SQCC. 
Due to the nature of the quantum-classical signaling scheme (where a small zero-mean quantum modulation is applied onto a large classically-modulated coherent state), the implementation of SQCC ideally requires the full CSI at Bob in order for him to precisely displace the received state (with a non-zero mean amplitude) back to its zero-mean version (either physically or virtually) under every channel realization. 
When the full CSI is unavailable at Bob, however, the imperfect displacement using the channel statistics (rather than the full CSI) will introduce additional noise to the decoding of quantum information. 
Although this scenario (i.e., Bob does not have the full CSI) is certainly not ideal for SQCC, we expect that the use of diversity should reduce the additional noise induced by the imperfect displacement since independent subchannel transmission, together with diversity combining, effectively reduces the fluctuation of the overall transmissivity around its mean. 
In other words, we expect the use of diversity to enhance the performance of SQCC over FSO fading channels under a practical scenario where the full CSI is unavailable at Bob. 

{{For a fair comparison across scenarios where different numbers of subchannels are used, we have assumed a fixed size for all receiving apertures, ensuring that the total received signal power at the satellite remains statistically the same regardless of the number of subchannels used.
In addition to employing diversity (the effectiveness of which has been confirmed in this work), other methods for fading mitigation may also be beneficial in Earth-to-satellite CV quantum communications. The most intuitive alternative is the use of a single larger receiving aperture (despite the practical infeasibility of aperture averaging at the satellite). 
Considering the limited space available for a satellite, future research could involve comparing a single large receiving aperture with multiple smaller receiving apertures (occupying the same total area as the large receiving aperture) on the satellite and investigating how many smaller apertures (and of what size) are required to outperform a single large aperture.}}

{{It is worth noting that although the large coherence length at the satellite effectively negates the practical usefulness of engineering receiver-aperture separation at a single satellite, it is possible to exploit an additional degree of diversity and/or to achieve aperture averaging via the use of a distributed satellite constellation (where satellites are separated by large distances).
Within this context, another useful future study may involve a detailed investigation into the formation and coordination of a satellite constellation for exploiting such diversity via appropriate inter-satellite distances.}}

\section{Conclusions}\label{Sec:Conclusions}
In this work, we explored the feasibility and effectiveness of utilizing spatial diversity as a means of fading mitigation in Earth-to-satellite CV quantum communications. 
We first determined a practical system model where spatial diversity can potentially be introduced to improve system performance within the context of satellite-based CV quantum communications. Based on our system model, we proposed a practical diversity-assisted scheme for the tasks of entanglement distribution and coherent-state transfer, and we analyzed the evolution of quantum states within these tasks. 
Using our numerical simulations for the realistic modelling of Earth-to-satellite channels, we perform a detailed performance investigation on diversity-assisted entanglement distribution, confirming the usefulness of diversity in Earth-to-satellite quantum communications. 
Within the context of SQCC over fading channels, our further results indicate that the transfer of both large (classically-modulated) coherent states (for the classical part of communications) and small (quantum-modulated) coherent states (for the quantum part of communications) can benefit from the use of diversity. 
The usefulness of the uplink configuration in space-based quantum communications is further elaborated on in Sec.~\ref{Sec:UplinkUsefulness} of~\cite{SM}.
Note that our results can be applied to arbitrary FSO fading channels.

This work provides valuable insights into the practical implementation of entanglement distribution and coherent-state transfer within the context of satellite-based CV quantum communications, filling the largely unresolved gap towards the realization of the global-scale quantum Internet.

\section{Acknowledgment}
The Australian Government supported this research through the Australian Research Council's Linkage Projects funding scheme (Project No. LP200100601). The views expressed herein are those of the authors and are not necessarily those of the Australian Government or the Australian Research Council. Approved for Public Release: 24-1118.

\appendix
\renewcommand\thesection{\Alph{section}}
\renewcommand\thesubsection{\arabic{subsection}}
\renewcommand{\theequation}{A\arabic{equation}}
\renewcommand{\theHequation}{A\arabic{equation}}
\renewcommand{\thefigure}{A\arabic{figure}}   
\renewcommand{\theHfigure}{A\arabic{figure}} 
\renewcommand{\thetable}{A\arabic{table}}  
\renewcommand{\theHtable}{A\arabic{table}}  
\setcounter{figure}{0}   
\setcounter{equation}{0}

\section{Channel Modeling}\label{Sec:ChannelModelling}
\subsection{Transmissivity Fluctuation within Atmospheric Channels}\label{Sec:AtmosProp}
The fluctuating subchannel transmissivities $\{T_j\}$ are random variables whose probability distributions are determined by the small random refractive index fluctuations (i.e., optical turbulence) caused by the random inhomogeneities (i.e., turbulent eddies) within the Earth's turbulent atmosphere. 
These random inhomogeneities cause continuous phase modulations, imposing amplitude and phase distortions on an optical beam as it propagates through an atmospheric channel. 
Particularly, the turbulence-induced effects (e.g., beam wandering, beam-shape deformation, beam broadening, and intensity scintillation) related to amplitude distortions can negatively affect the performance an FSO communication system by introducing substantial fluctuating losses in real-world scenarios where a finite-sized aperture is used at the receiver~\cite{book}. 
Under the paraxial approximation, the propagation of a monochromatic optical beam $\psi(\mathbf{R})$ through the turbulent atmosphere is governed by the stochastic parabolic equation
\begin{equation}\label{Eq.sHe}
	\nabla_{\text{T}}^{2} \psi(\mathbf{R})+i 2 k \frac{\partial\psi(\mathbf{R})}{\partial {z}} +2 \delta n(\mathbf{R}) k^{2} \psi(\mathbf{R})=0,
\end{equation}
where $\mathbf{R}=[x,y,z]^{\top}$ is the three-dimensional position vector, $n(\mathbf{R})$ denotes the the refractive index at $\mathbf{R}$, $\delta n(\mathbf{R})\!=\!n(\mathbf{R}) - \langle n(\mathbf{R})\rangle$ denotes the small refractive index fluctuations. and  $\nabla_{\text{T}}^{2}=\partial^{2} / \partial x^{2}+\partial^{2} / \partial y^{2}$ is the transverse Laplacian operator~\cite{book}. Note that, Eq.~(\ref{Eq.sHe}) is implicitly based on the commonly used assumption for atmospheric turbulence 
\begin{equation}\label{Eq.OpticalTurbulenceAssumptions}
\langle n(\mathbf{R}) \rangle\!=\!1, \,\, \delta n(\mathbf{R})\!\ll\!1, \,\, \ensavg{\delta n(\mathbf{R})}=0.
\end{equation}
In this work, we assume that all the transmitted optical beams are in the fundamental transverse Gaussian mode. The general form of a Gaussian beam propagating along the $z$ axis is give by
\begin{equation}\label{Eq.Gaussian}
	\begin{aligned} \psi_{\text{G}}(\mathbf{r},z)&=\sqrt{\frac{2}{\pi}} \frac { 1 } { w ( z ) }  \exp \left[ \frac { - |\mathbf{r}| ^ { 2 } } { w ^ { 2 } ( z ) } \right]\exp \left[i\phi(\mathbf{r},z)\right],
	\end{aligned}
\end{equation}
where $\mathbf{r}=[x,y]^{\top}$ denotes the two-dimensional position vector in the transverse plane, $w(z)=w_{0} \sqrt{1+(z / z_{\text{R}})^{2}}$, $w_0$ is the beam-waist radius, $z_{\mathrm{R}}=\pi w_{0}^{2}/\lambda$ is the Rayleigh range, $\lambda$ is the (central) optical wavelength, and $\phi(\mathbf{r},z) =  \frac {  k |\mathbf{r}| ^ { 2 } z } { 2 \left( z ^ { 2 } + z _ { \text{R} } ^ { 2 } \right) }  - \arctan \left( \frac { z } { z _ { \text{R} } } \right)$ with $k=2\pi/\lambda$ being the optical wavenumber~\cite{SQCC_Matt2024}. 

Denoting the transmitted beam as $\psi_{\text{G}}(\mathbf{r},0)$  and the received beam as $\psi_{\text{Rx}}(\mathbf{r},L)$, we can express the (fluctuating) channel transmissivity as 
\begin{equation}\label{Eq.eta}
	T = \int_{\mathcal{A}} \left|\psi_{\text{Rx}}(\mathbf{r},L)\right |^2\,\text{d}^2\mathbf{r},
\end{equation}
where $L$ denotes the channel distance, and $\mathcal{A}$ denotes the aperture area at the receiver. We assume a circular receiver aperture whose radius is $r_{\text{a}}$. The corresponding channel loss is simply given by $\text{Loss [dB]}=-10\log_{10}T$.

In this work, we adopt the commonly-used metrics to characterize the fluctuating channel loss -- we quantify the \textit{mean channel loss} using the average value (i.e., expectation in dB) of the fluctuating loss and the \textit{channel fading strength} using the standard deviation (in dB) of the fluctuating loss in any given subchannel.

For simplicity, when describing our calculations, we formally arrange the subchannel transmissivities in a vector $\mathbf{T} = [T_1,\,T_2,\,\dots,\,T_M]^{\top}$, where $T_{j}$ denotes the random transmissivity of subchannel $j$ ($1\le j \le M$). We use $P_{\text{chn}}(\mathbf{T})$ to denote joint Probability Density Function (PDF) of subchannel transmissivities. Averaging over fluctuating subchannel transmissivities $\{T_j\}$ is carried out via Monte Carlo simulations. When the phase-screen simulations (to be introduced in Appendix~\ref{Sec:PhsDetails} and Appendix~\ref{Sec:SimulationSettings}) and the log-normal model (to be introduced in Appendix~\ref{Sec:LNDetails}) are used for channel modelling, we generate 1000 and 3000 independent realizations for each one of the $M$ used subchannels, respectively.

\subsection{Phase-Screen Simulations for Realistic Uplink Channels}\label{Sec:PhsDetails}
In order to faithfully capture the effects imposed by the optical turbulence on a propagating optical beam within the Earth’s atmosphere, we investigate the atmospheric propagation of optical beams by numerically solving Eq.~(\ref{Eq.sHe}) using the split-step method~\cite{SimAP}. The practical implementation of the split-step method is commonly referred to as phase-screen simulations, where the atmospheric channel is modelled as a set of slabs with a random phase screen located in the midway of each slab. Two vacuum propagations with one random phase modulation in between are repeatedly performed until the beam reaches the receiver. 
The implementation of our numerical simulations in this work largely follows our previous work~\cite{EduardoEnhanced2021}. 
Specifically, the implementation of our phase screen simulations requires careful configuration and optimization based on the evaluation of the scintillation index (which is the normalized variance of intensity) and the Fried parameter (which quantifies the coherence length of turbulence-induced phase errors in the transverse plane) for both the whole channel and each individual slab. 
More technical details regarding the implementation of our numerical simulations can be found in our previous works~\cite{Ziqing_OAMQKD,Ziqing_TMQKD,EduardoEnhanced2021,SQCC_Matt2024} and the references therein.

{{Despite being computationally intensive, our phase-screen simulations represent one of the most powerful and flexible methods to investigate optical propagation within the atmosphere of the Earth. 
Although our simulations do not yield analytical results, they implicitly capture all the well-known turbulence-induced effects (including scintillation, beam wandering, beam deformation, and beam broadening) without relying on the approximations (e.g., those related to certain turbulence parameters and the final beam shape) commonly employed by state-of-the-art theoretical studies (e.g.,~\cite{EllipticBeam2016,SatMedLink,PirandolaSatellite2021,PirandolaFSO2022,PirandolaSatellite2024}).
Our simulations have been extensively applied within the context of terrestrial and satellite-based optical quantum communications (e.g.,~\cite{SQCC_Matt2024,EduardoEnhanced2021,Ziqing_OAMQKD,Ziqing_TMQKD,EduardoDSTData2020}), with their accuracy confirmed through real-world experiments~\cite{EduardoDSTData2020} and cross-validated in~\cite{EduardoEnhanced2021} against state-of-the-art analytical models, particularly the family of models based on the elliptic-beam approximation (originally proposed in~\cite{EllipticBeam2016}). 
We refer the reader to~\cite{EduardoEnhanced2021} for a review of the literature covering other state-of-the-art channel modeling techniques.}}

For illustration purposes, in Fig.~\ref{fig:PDF_Loss} we adopt the simulation settings in Appendix~\ref{Sec:SimulationSettings} and plot the PDF of the fluctuating channel loss, achieved under different zenith angles (denoted as $\theta_{\text{z}}$), over a realistic Earth-to-satellite channel. Some important loss statistics acquired from the simulation results on channel loss presented in Fig.~\ref{fig:PDF_Loss} are further explicitly given in Table~\ref{Table:LossStats}.
\begin{figure}[htbp!]
\hspace{0.4em}\includegraphics[width=1.05\linewidth]{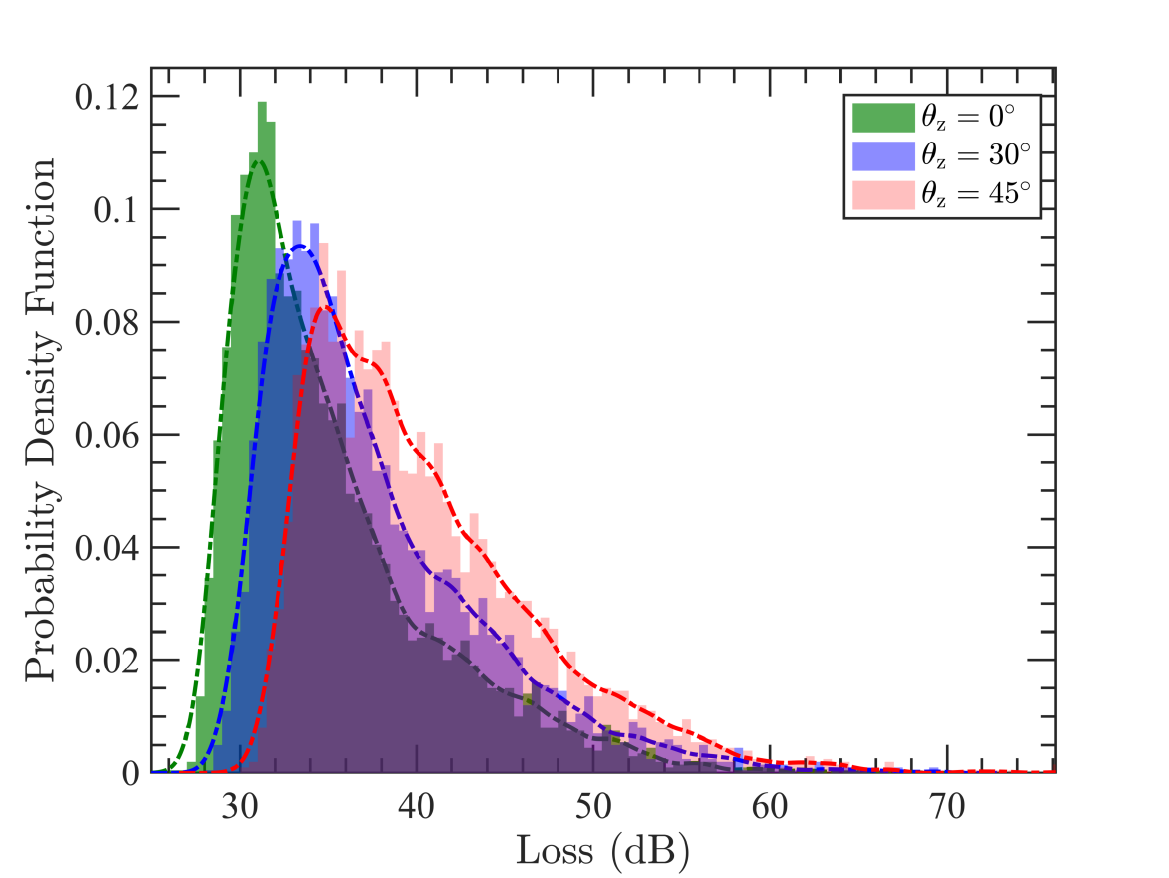}
\caption{Simulated PDF of channel loss under different $\theta_{\text{z}}$ values over an Earth-to-satellite channel. The PDF curves are generated by fitting a kernel probability distribution to our discrete simulation data (semi-transparent) on channel loss using the \textit{fitdist} function provided by {\sc{Matlab}}. The simulation settings are presented in Appendix~\ref{Sec:SimulationSettings}.}\label{fig:PDF_Loss}
\end{figure}

\begin{table*}[bth!]
\centering
\setlength{\tabcolsep}{14pt}
\renewcommand{\arraystretch}{1.5}
\caption{Loss statistics acquired from the simulation results on channel loss presented in Fig.~\ref{fig:PDF_Loss}.}\label{Table:LossStats}
\begin{tabular}{@{}ccccc@{}}
\toprule
 & Minimum Loss & \begin{tabular}[c]{@{}c@{}}Diffraction Loss \\ (If No Turbulence)\end{tabular} & Mean Loss / Fluctuation Strength & Maximum Loss~ \\ \midrule
$\theta_{\text{z}}=0^{\circ}$  & 26.5 dB & 27.2 dB & 35.2 dB / 5.8 dB & 67.8 dB \\
$\theta_{\text{z}}=30^{\circ}$ & 28.5 dB & 28.4 dB & 37.6 dB / 6.2 dB & 69.4 dB \\
$\theta_{\text{z}}=45^{\circ}$ & 30.1 dB & 30.2 dB & 40.4 dB / 6.4 dB & 76.2 dB \\ \bottomrule
\end{tabular}
\end{table*}

\subsection{Simulation Settings}\label{Sec:SimulationSettings}\label{Sec:NS}
The real-world uplink settings for our phase-screen simulations are detailed as follows.
We restrict ourselves to the case of an LEO satellite with a satellite altitude $H=500\,\text{km}$, and we consider zenith angles $\theta_{\text{z}}$ ranging from $0^{\circ}$ to $45^{\circ}$. For simplicity, we assume the satellite is at the same zenith angle from the perspective of all the transmitting ground stations. 
For the atmospheric parameters, we follow our previous works (e.g.,~\cite{Ziqing_OAMQKD,Ziqing_TMQKD,SQCC_Matt2024}) and set the ground-level turbulence strength to $A=9.6\times 10^{-14}\, \text{m}^{-2 / 3}$, the outer scale to $L_{\text{outer}} = 5\,\text{m}$, and the inner scale to $l_{\text{inner}}=1\,\text{cm}$. 
For the wind speed profile, we adopt the Bufton wind model and set the ground-level wind speed to $V_{\text{g}}=3\,\text{m/s}$, giving a root-mean-squared wind speed $v_{\text{rms}}=21\,\text{m/s}$ (see~\cite{Ziqing_OAMQKD} for detailed explanations of these parameters). 
For the optical parameters, we assume all the ground stations transmit at $\lambda=1064\,\text{nm}$. In order to determine the feasibility of our scheme utilizing the existing operational platforms, we set the beam-waist radius (receiver aperture radius) to $w_0=3.5\,\text{cm}$ ($r_a=15\,\text{cm}$) to mimic the setting used in the recent demonstration of Earth-to-satellite quantum teleportation via the Micius quantum satellite~\cite{UplinkTeleportation2017}.

\subsection{Log-Normal Model for Channel Loss}\label{Sec:LNDetails}
In order to study the general behaviours of our diversity-assisted system over fluctuating channels, sometimes it is also useful to adopt a simple flexible model (where the channel loss follow a simple but not necessarily realistic probability distribution) to describe the  fluctuating channel loss. 
Particularly, it is desirable that the channel-loss distribution used to construct the model
\begin{enumerate}
    \item[1)] has support $[0,+\infty)$,
    \item[2)] allows for the independent and flexible setting of mean and variance, 
    \item[3)] can be determined by as few parameters as possible, and 
    \item[4)] is not derived by truncation or combination of other distributions.
\end{enumerate}

In this work, we choose the log-normal distribution (which is a continuous probability distribution of a random variable whose logarithm is normally distributed) to construct our simple and flexible model since this distribution satisfies all the aforementioned features. 
In this model, the random channel loss is modelled as a random variable following log-normal distribution  ${\text{Loss}}_{\text{LgN}} [\text{dB}]\sim \text{LgN}\,(\mu_{\text{LgN}},\sigma^{2}_{\text{LgN}})$ with mean value $\mu_{\text{LgN}}\,[\text{dB}]$ (which quantifies the mean channel loss) and standard deviation  $\sigma_{\text{LgN}}\,[\text{dB}]$ (which quantifies the channel fading strength). 
The realizations of such a log-normal random variable can be readily generated using the following property
\begin{equation}
   \operatorname{ln}({\text{Loss}}_{\text{LgN}}) \sim \mathcal{N} \left(\ln \left(\frac{\mu_{\text{LgN}}^2}{\sqrt{\mu_{\text{LgN}}^2+\sigma_{\text{LgN}}^2}}\right),\ln \left(1+\frac{\sigma_{\text{LgN}}^2}{\mu_{\text{LgN}}^2}\right)\right),
\end{equation}
where $\mathcal{N}(\cdot,\cdot)$ denotes normal distribution with the first (second) argument being the mean (variance).

\section{Beam Splitter Transformation}\label{Sec:BS}
\subsection{General Description}\label{Sec:BS_General}
\begin{figure}[hbt!]
\centering
\includegraphics[width=0.6\linewidth]{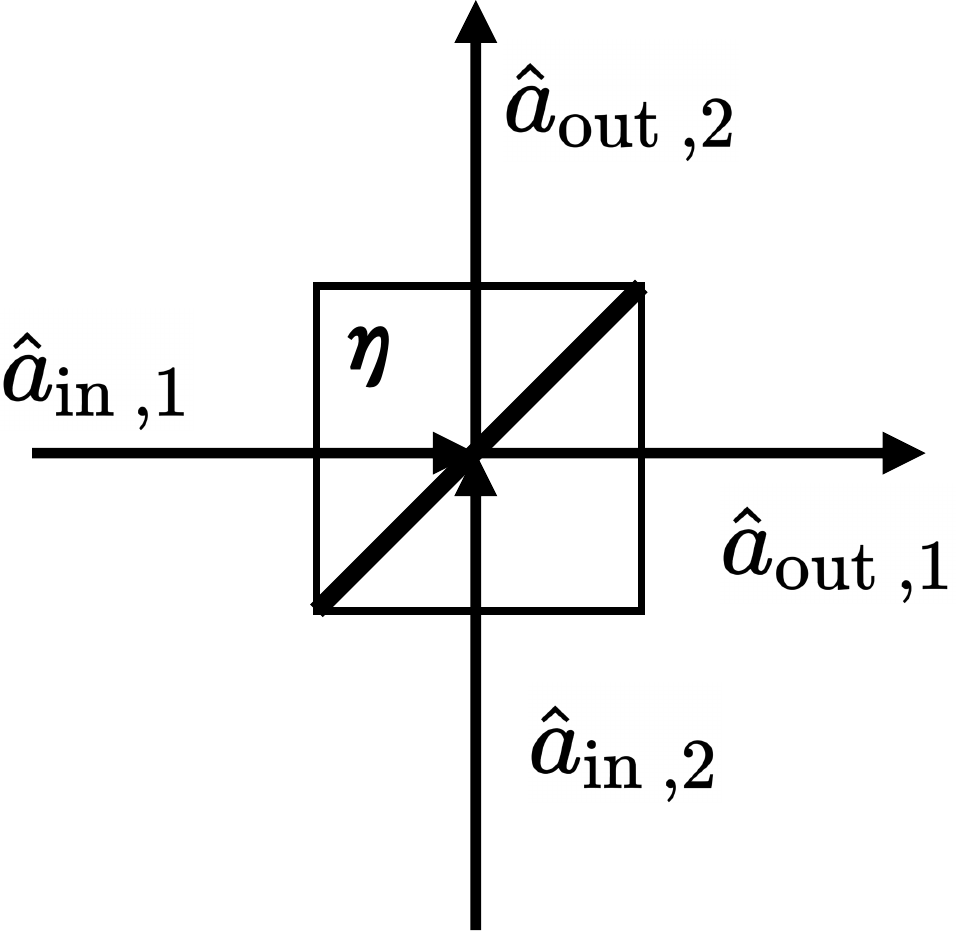}
\caption{The beam splitter model adopted in this work.}\label{Fig.BS}
\end{figure}

\setcounter{equation}{5}
As a Gaussian transformation that transforms Gaussian states into Gaussian states, the beam splitter transformation is one of the most important linear optics transformations for two bosonic modes.
The convention of the beam splitter transformation adopted in this work is underpinned by the beam splitter model illustrated in Fig.~\ref{Fig.BS}.
Specifically, a beam splitter with a real transmission coefficient $\eta$ ($0\le \eta\le 1$, $\eta\in\mathbb{R}$) performs the following linear mode transformation in the Heisenberg picture
\begin{equation}\label{Eq:SingleBS}
\renewcommand\arraystretch{1.8}  
\begin{bmatrix}
\hat{a}_\text{out,1} \\
\hat{a}_\text{out,2}
\end{bmatrix}
=
\begin{bmatrix}
\eta                    & \sqrt{1-{\eta}^2}\\
\sqrt{1-{\eta}^2}     & -\eta
\end{bmatrix}
 \begin{bmatrix}
\hat{a}_\text{in,1}\\
\hat{a}_\text{in,2}
\end{bmatrix},
\end{equation}
where $\hat{a}_\text{in,1}$, $\hat{a}_\text{in,2}$, $\hat{a}_\text{out,1}$, and $\hat{a}_\text{out,2}$ denote the annihilation operators for input mode 1, input mode 2, output  mode 1 (i.e., the transmitted mode), and output mode 2 (i.e., the reflected mode), respectively. 
Note that we reserve the use of the symbol $\eta$ (without any subscript) exclusively for the transmission coefficient in the general description of the beam splitter transformation.

Under the CF formalism, the beam splitter transformation in Eq.~(\ref{Eq:SingleBS}) corresponds to a CF argument transformation that leads to the following result
\begin{equation}\label{Eq:BS_Out}
\chi_{\text{out}}(\xi_1,\xi_2)=\chi_{\text{in}}\left(\eta \xi_1+\sqrt{1-{\eta}^2}\xi_2,\sqrt{1-{\eta}^2}\xi_1-\eta\xi_2\right),
\end{equation}
where $\chi_{\text{in}}(\cdot,\cdot)$ ($\chi_{\text{out}}(\cdot,\cdot)$) denotes the CF of the two-mode input (output) state -- see, e.g.,~\cite{RealisticCVTeleportationNG2010,CF_Paper2013}.
In fact, the result of any linear optics transformation can be expressed as a transformation of CF arguments under the CF formalism. 
The CF formalism also provides a convenient way to describe a reduced quantum state -- tracing out one mode can be simply achieved by setting the corresponding CF argument to zero~\cite{CVBook}. For example, after tracing out the output mode 2 of the aforementioned beam splitter, the CF of the reduced state of output mode 1 can be simply given by
\begin{equation}\label{Eq:BS_Out_PT}
\chi_{\text{out},1}(\xi_1) = \chi_{\text{out}}(\xi_1,0) =\chi_{\text{in}}\left(\eta \xi_1,\sqrt{1-{\eta}^2}\xi_1\right).
\end{equation}

The statistical-moment description can also be conveniently used to describe a beam splitter transformation. Specifically, for the linear beam-splitter transformation given in Eq.~(\ref{Eq:SingleBS}), the quadrature operators $\hat{\mathbf{x}}_{\text{in}}=\left[\hat{q}_{\text{in,1}}, \hat{p}_{\text{in,1}}, \hat{q}_{\text{in,2}}, \hat{p}_{\text{in,2}}\right]^{\top}$ are transformed via the symplectic map
\begin{equation}
    \hat{\mathbf{x}}_{\text{out}} = \mathbf{B}(\eta) \hat{\mathbf{x}}_{\text{in}}, 
\end{equation}
where $\mathbf{B}(\eta)=\left[\begin{array}{cc}
\eta \mathbb{1}_2 & \sqrt{1-\eta^2} \mathbb{1}_2 \\
\sqrt{1-\eta^2} \mathbb{1}_2 & -\eta \mathbb{1}_2
\end{array}\right]$ is the corresponding symplectic matrix satisfying $\mathbf{B}(\eta) \boldsymbol{\Omega} \mathbf{B}(\eta)^{\top}=\boldsymbol{\Omega}$.
As a result, in terms of the statistical moments $\bar{\mathbf{x}}$ and $\mathbf{V}$, the action of the beam splitter is characterized by the following transformations
\begin{equation}\label{Eq:MomentTransBS}
    \bar{\mathbf{x}}_{\text{out}} = \mathbf{B}(\eta) \bar{\mathbf{x}}_{\text{in}}, \quad \mathbf{V}_{\text{out}} = \mathbf{B}(\eta) \mathbf{V}_{\text{in}} \mathbf{B}(\eta)^{\top}.
\end{equation}
Under the statistical-moment description, tracing out one mode can be simply achieved by discarding the corresponding elements in the first-order statistical moment $\bar{\mathbf{x}}$ and the corresponding rows and columns in the CM $\mathbf{V}$.

\subsection{Beam-Splitter Description of Atmospheric Channel}\label{Sec:BS_Channel}
The beam splitter transformation is ubiquitously used in this work.
Specifically, we follow the common practice and model an atmospheric quantum channel as a lossy channel with a real transmissivity $T$ and a total excess noise $\epsilon$ (see e.g.,~\cite{GaussianQuantumInformation,TheoryOfPracticalImplementation}). Such a channel preserves the Gaussian characters of a quantum state and can be conveniently modelled as a beam splitter (with transmission coefficient $\sqrt{T}$) in a thermal background with a certain mean number of photons. 
In this work, we explicitly describe an atmospheric quantum channel as a beam splitter with input mode $\hat{a}_{\text{in,1}}$ being the quantum signal and input mode $\hat{a}_{\text{in,2}}$ being the associated environmental mode in a thermal state with equal quadrature variance $V_{\text{thermal}}=1+\frac{\epsilon}{1-T}$~\cite{TheoryOfPracticalImplementation}. 

\subsection{Beam-Splitter Description of Diversity Combining}\label{Sec:BS_Combine}
In this work, the combining of optical signals is assumed to be achieved using beam splitters. Specifically, we always assume that only the transmitted mode (i.e., $\hat{a}_\text{out,1}$) of a combining beam splitter is used as the output of the combining operation -- this means the unused output port (i.e., $\hat{a}_\text{out,2}$) should always be traced out.

\section{Diversity Combining}\label{Sec:DiversityCombining}
In an $M\times M$ diversity scheme using $M$ subchannels, at least $M-1$ beam splitters are required in a combining module to combine the $M$ incoming signals. 
We denote the $M$ incoming quantum states to be combined as $\smash{\{\rho_j\}}$ ($\smash{1 \le j \le M}$), the transmission coefficient of the $l^{\text{th}}$ combining beam splitter ($\smash{1\le l \le M-1}$) as $\eta_l$ ($\smash{0\le\eta_l\le1,\,\,\eta_l\in\mathbb{R}}$), and the resulting quantum state from the combining of the $M$ quantum states as $\rho_{\text{DC}}^{(M)}$.

\begin{figure*}[bt!]
\centering
\includegraphics[width=0.8\linewidth]{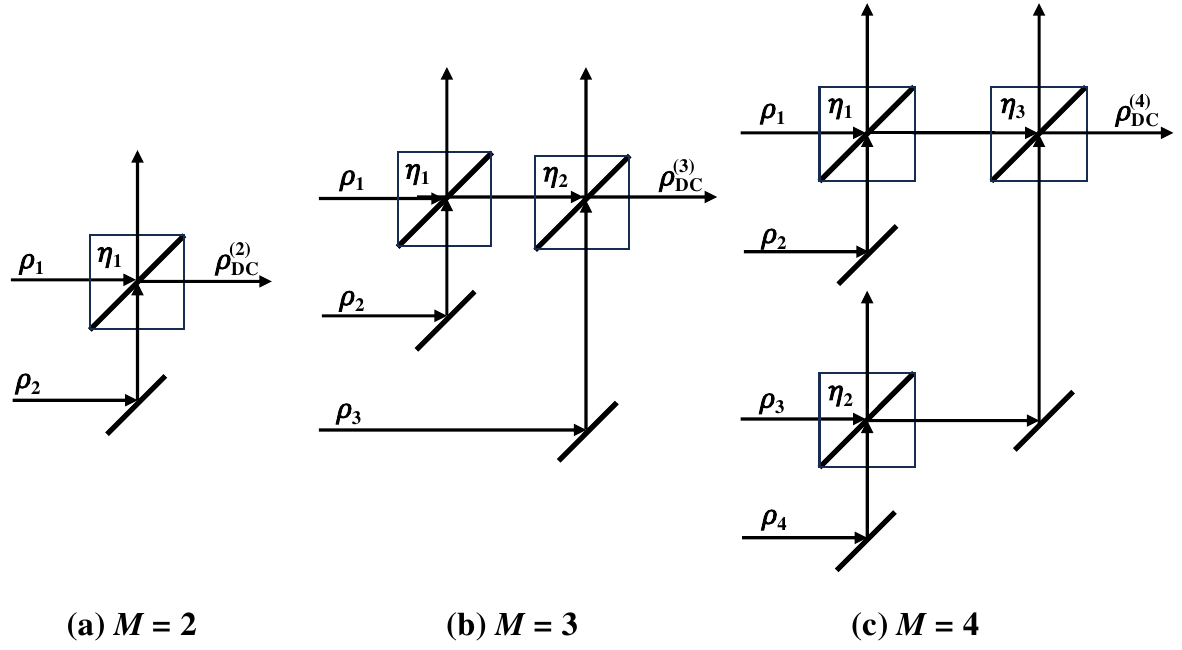}
\caption{Configuration of beam splitters in the combining module. The unlabeled output ports are ignored since the outputs from those ports are discarded. Note that, the arrangement of beam splitters in a combining module is not unique.}\label{Fig.Combining}
\end{figure*}

In Fig.~\ref{Fig.Combining}, we illustrate the configuration of beam splitters in our combining module for $M=1,\,2,\,3,\,\text{and}\,\,4$.
In this work, we set the combining module to coherently combine each subchannel signal with an equal weight of $1/\sqrt{M}$ due to the lack of knowledge of instantaneous subchannel transmissivities and subchannel noises. 
Considering the equal mean subchannel powers resulting from the equal mean subchannel transmissivities under the assumption of i.i.d. subchannels, one can readily show that $1/\sqrt{M}$ is the statistically optimal combining weight achieved by setting $\{\eta_l\}$ such that zero signal power comes out of the unused output port of every beam splitter within the combining module.
Given an equal combining weight of $1/\sqrt{M}$, the transmission coefficients $\{\eta_l\}$ of the combining beam splitters can be determined based on the specific arrangement in Fig.~\ref{Fig.Combining} and the convention we use (see Fig.~\ref{Fig.BS} and Eq.~(\ref{Eq:SingleBS})).
For example, in Fig.~\ref{Fig.Combining}(b) such a setting corresponds to $\eta_1=1/\sqrt{2}$ and $\eta_2=\sqrt{2/3}$, leading to the combining of  $\rho_1$, $\rho_2$, and $\rho_3$ with an equal weight $1/\sqrt{3}$. 

It should be noted that, the above configuration does not assume any specific type of input quantum states and can thus be directly used for diversity combining in any quantum information protocols.  For example, in entanglement distribution, $\rho_j$ represents the $j^{\text{th}}$ part of mode B (of Alice's TMSV state $\rho_{\text{AB}}$), which is transmitted through subchannel~$j$. 
In coherent-state transfer, $\rho_j$ represents the $j^{\text{th}}$ part (of Alice's encoded coherent state $\ket{\alpha_{\text{tar}}}$), which is transmitted through subchannel~$j$.

Here we elaborate on the real-world practicality of using the combining module, which consists mainly of beam splitters (note that auxiliary phase shifters may be needed in practice). The beam splitter is one of the simplest optical devices required in every quantum or classical optical communications system. 
In the standard practical implementation of free-space optical communications where the signal beam illuminating a tracked receiving aperture is first coupled into a fiber, the combining module can be constructed using multiple fiber-based beam splitters, which are virtually identical to their free-space counterparts. 
The combing module can also be integrated entirely onto a photonic chip (see an example of a 15-subchannel coherent combiner for FSO communications in~\cite{CombiningModuleChip2021}), further simplifying the combining operation.

\begin{widetext}
\section{Analytical Expression of Fidelity}\label{A:AnalyticalFidelity}
\setcounter{equation}{10}
Denoting the encoded coherent state as $\ket{\alpha_{\text{tar}}}$ and the transmitted coherent state (same for all the transmitters due to the equal splitting) as $\ket{\alpha^{(M)}_{\text{tx}}}=\ket{\frac{\alpha_{\text{tar}}}{\sqrt{M}}}$, we can analytically express the fidelity (before being averaged over the quantum alphabet and the fluctuating subchannel losses) in Eq.~(\ref{Eq.AvgFidelity}) of the main text as 
\begin{equation}\label{Eq.SingleFidelityAnalytical}
\mathcal{F}\left(\rho_{\text{DC}}^{(M)}, \ket{\alpha_{\text{tar}}^{\prime}}\bra{\alpha_{\text{tar}}^{\prime}}\right)=\frac{2}{Y_{(M)}} \exp \left\{-\frac{2 \left[X_{(M)}^2 \left|\alpha_{\text{tx}}^{(M)}\right|^2-2X_{(M)} \big[\mathfrak{Re}({\alpha_{\text{tx}}^{(M)}}) \mathfrak{Re}({\alpha_{\text{tar}}^{\prime}}) + \mathfrak{Im}({\alpha_{\text{tx}}^{(M)}})\mathfrak{Im}({\alpha_{\text{tar}}^{\prime}})\big]+\left|\alpha_{\text{tar}}^{\prime}\right|^2\right]}{Y_{(M)}}\right\},
\end{equation}
where $\ket{\alpha_{\text{tar}}^{\prime}}=\ket{{\alpha_{\text{tar}}}/{\langle \sqrt{T} \rangle}}$ denotes the scaled encoded state (recall Sec.~\ref{Sec:MetricCT} of the main text). In Eq.~(\ref{Eq.SingleFidelityAnalytical}), $X_{(M)}$ and $Y_{(M)}$ are given by
\begin{equation}\label{Eq:XM}
X_{(M)}=
    \begin{cases}
    \sqrt{T_1},\,\,& M=1\\
    \eta_1 \sqrt{T_1} + \sqrt{1-\eta_1^2} \sqrt{T_2},\,\,    & M=2\\
  \eta_2\left( \eta_1 \sqrt{T_{1}} + \sqrt{(1-\eta_1^2) T_{2}} \right) + \sqrt{(1-\eta_2^2) T_{3}},\,\, & M=3\\
   \eta_3\left(\eta_1 \sqrt{T_{1}}+\sqrt{(1-\eta_1^2) T_{2}}\right)+\eta_2 \sqrt{(1-\eta_3^2) T_{3}}+\sqrt{(1-\eta_2^2)(1-\eta_3^2) T_{4}},\,\,   & M=4
\end{cases},
\end{equation}

\begin{equation}\label{Eq:YM}
Y_{(M)}=
    \begin{cases}
   \epsilon_1+2,& M=1\\
    \left[\eta_1^2\left(\epsilon_1-\epsilon_2\right)+\epsilon_2\right]+2,    & M=2\\
    \left[\eta_2^2\big(\eta_1^2\left(\epsilon_1-\epsilon_2\right)+\epsilon_2-\epsilon_3\big)+\epsilon_3\right]+2, & M=3\\
  \left[\eta_1^2 \eta_3^2\left(\epsilon_1-\epsilon_2\right)+\epsilon_4+\eta_2^2\left(\epsilon_3-\epsilon_4\right)+\eta_3^2\big(\epsilon_2-\eta_2^2\left(\epsilon_3-\epsilon_4\right)+\epsilon_4\big)\right]+2,   & M=4
\end{cases},
\end{equation}
where $\{T_j\}$ denote the subchannel transmissivities ($1\le j\le M$), $\{\epsilon_j\}$ denote the subchannel excess noises ($1\le j\le M$), and $\{\eta_l\}$ denote the transmission coefficients of the combining beam splitters ($1\le l\le M-1$). 
For generality, in Eqs.~(\ref{Eq:XM})(\ref{Eq:YM}) we have chosen not to plug in any specific values for $\{\eta_l\}$.
\end{widetext}

\input{Main_v2.bbl}
\input{SM_v2}

\input{SM_v2.bbl}
\end{document}

%% file: SM_v2.tex
\onecolumngrid
\pagebreak

\begin{center}
{\Large Supplementary Material for}\\
~\\
{\large \bf{Exploiting Spatial Diversity in Earth-to-Satellite Quantum-Classical Communications}}\label{Sec:Background}
\end{center}
\normalsize
\begin{center}
{Ziqing Wang$^1$, Timothy~C.~Ralph$^2$, Ryan~Aguinaldo$^{3}$ and~Robert~Malaney$^1$}\\
\vspace{4pt}
\textit{$^{1}$School of Electrical Engineering and Telecommunications,\\ University of New South Wales, Sydney, NSW 2052, Australia}\\
\textit{$^{2}$Centre for Quantum Computation and Communication Technology,\\
School of Mathematics and Physics, University of Queensland, St Lucia, QLD 4072, Australia}\\
\textit{$^{3}$Northrop Grumman Corporation, San Diego, CA 92128, USA}\\
 \vspace{1em}
 (Dated: \today)
\end{center}
\renewcommand\thesection{S-\Roman{section}}
\renewcommand\thesubsection{\Alph{subsection}}
\setcounter{figure}{0}   
\setcounter{equation}{0}
\renewcommand{\theequation}{S\arabic{equation}}
\renewcommand{\theHequation}{S\arabic{equation}}
\renewcommand{\thefigure}{S\arabic{figure}}   
\renewcommand{\theHfigure}{S\arabic{figure}} 
\renewcommand{\thetable}{S\arabic{table}}  
\renewcommand{\theHtable}{S\arabic{table}}  
\renewcommand{\appendixname}{Section}  
\setcounter{section}{0}   
\setcounter{page}{1}  
\renewcommand{\thepage}{Supplementary Material Page~\arabic{page}}
\renewcommand{\bibnumfmt}[1]{[S#1]}
\renewcommand{\citenumfont}[1]{S#1}

\bigskip
In this Supplementary Material, we briefly review the necessary background on the underlying formalism of this work. Most of the descriptions in this section are adapted from~\cite{GaussianQuantumInformation_SM,TheoryOfPracticalImplementation_SM}.

\section{Description of Bosonic Systems}\label{Sec:DBS}
For an $N$-mode bosonic system consisting of individual bosonic modes with mode operator pairs $\{\hat{a}_n, \hat{a}_n^{\dagger}\}_{n=1}^N$ (where $\hat{a}_n$ is the annihilation operator and $\hat{a}_n^{\dagger}$ is the creation operator for mode $n$), a pair of quadrature operators $\left\{\hat{q}_n=\hat{a}_n+\hat{a}_n^{\dagger}, \hat{p}_n=i(\hat{a}_n-\hat{a}^{\dagger}_n)\right\}$ can be defined for each consisting mode. 
After conveniently arranging the field operators into a vectorial operator $\hat{\mathbf{b}}:=[\hat{a}_1, \hat{a}_1^{\dagger}, \ldots, \hat{a}_N, \hat{a}_N^{\dagger}]^{\top}$ and the quadrature operators into another vectorial operator $\hat{\mathbf{x}}:=[\hat{q}_1, \hat{p}_1, \ldots, \hat{q}_N, \hat{p}_N]^{\top}$, the bosonic commutation relations can be given by
\begin{equation}\label{Eq:FieldOperatorCCR}
    \left[\hat{b}_i, \hat{b}_j\right]=\Omega_{i j} \quad(i, j=1, \ldots, 2 N),
\end{equation}
and the canonical communication relation can be given by
\begin{equation}\label{Eq:QuadOperatorCCR}
    \left[\hat{x}_i, \hat{x}_j\right]=2 i \Omega_{i j}.
\end{equation}
In Eqs.~(\ref{Eq:FieldOperatorCCR})(\ref{Eq:QuadOperatorCCR}), $\Omega_{ij}$ denotes the generic element of the $2N \times 2N$ matrix
\begin{equation}\label{Eq:SymplecticForm}
\boldsymbol{\Omega}:=\bigoplus_{n=1}^N \boldsymbol{\omega}=\left[\begin{array}{ccc}
\boldsymbol{\omega} & & \\
& \ddots & \\
& & \boldsymbol{\omega}
\end{array}\right], \quad \boldsymbol{\omega}:=\left[\begin{array}{cc}
0 & 1 \\
-1 & 0
\end{array}\right],
\end{equation}
which is known as the the symplectic form. Note that we implicitly assume $\hbar=2$ throughout this work.

Every quantum state has an equivalent representation in terms of a quasi-probability distribution (i.e., the Wigner function) defined over a real symplectic space. 
Specifically, an arbitrary quantum state $\hat{\rho}$ is equivalent to a Wigner Characteristic Function (CF)
\begin{equation}
    \chi(\boldsymbol{\xi})=\operatorname{Tr}[\hat{\rho} D(\boldsymbol{\xi})],
\end{equation}
where $D(\boldsymbol{\xi}):=\exp \left(i \hat{\mathbf{x}}^{\top} \boldsymbol{\Omega} \boldsymbol{\xi}\right)$ is the Weyl operator, and $\boldsymbol{\xi} \in \mathbb{R}^{2 N}$. 
Via a Fourier transform, an arbitrary quantum state $\hat{\rho}$ can be shown to be equivalent to a Wigner function
\begin{equation}
    W(\mathbf{x})=\int_{\mathbb{R}^{2 N}} \frac{d^{2 N} \boldsymbol{\xi}}{(2 \pi)^{2 N}} \exp \left(-i \mathbf{x}^{\top} \boldsymbol{\Omega} \boldsymbol{\xi}\right) \chi(\boldsymbol{\xi}),
\end{equation}
where the continuous variables $\mathbf{x} \in \mathbb{R}^{2 N}$ are the eigenvalues (also referred to as the quadrature variables) of the quadrature operators $\hat{\mathbf{x}}$, spanning a real symplectic space $\left(\mathbb{R}^{2 N}, \boldsymbol{\Omega}\right)$ (i.e., the phase space).

A Gaussian state is a quantum state whose Wigner function is a Gaussian distribution of the quadrature variables. Such a state can be completely characterized by the first-order statistical moment and the second-order statistical moment of the quadrature operators -- the former is simply the mean value, and the latter is commonly presented in the form of a Covariance Matrix (CM).
The mean value (denoted as $\bar{\mathbf{x}}$) is given by
\begin{equation}
{\mathbf{\bar{x}}}:=\langle{\mathbf{\hat{x}}}\rangle=\operatorname{Tr}(\hat{\mathbf{x}} \hat{\rho}),
\end{equation}
and an arbitrary element of the CM (denoted as $\mathbf{V}$) is given by
\begin{equation}
    V_{i j}:=\frac{1}{2}\left\langle\left\{\Delta \hat{x}_i, \Delta \hat{x}_j\right\}\right\rangle,
\end{equation}
where $\Delta \hat{x}_i:=\hat{x}_i-\left\langle\hat{x}_i\right\rangle$ and $\{,\}$ denotes the anticommutator. The CM is a $2N\times 2N$ real and symmetric matrix that satisfies the uncertainty principle $\mathbf{V}+i \boldsymbol{\Omega} \geq 0$.

\section{Common Gaussian States}\label{Sec:GaussianStates}
\setcounter{equation}{7}
In this work we mainly work with Gaussian states.
The vacuum state $\ket{0}$ (given in the Fock basis) of a single mode is perhaps the most important Gaussian state. As its name suggests, a single-mode vacuum state has a mean photon number of zero (i.e., $\ensavg{\hat{a}^{\dagger}\hat{a}}=0$). 
The CF of a single-mode vacuum state is given by
\begin{equation}
    \chi_{\text{vac}}(\xi)=\exp{\left(-\frac{|\xi|^2}{2}\right)},
\end{equation}
where $\xi$ can be treated as a complex variable.
A single-mode vacuum state has zero mean and a CM equals to $\mathbb{1}_2=\operatorname{diag}(1,1)$.

A single-mode coherent state $\ket{\alpha}$ (with $\alpha \in \mathbb{C}$ being the complex amplitude) is generated by applying a displacement operator $D(\alpha)=\exp \left(\alpha \hat{a}^{\dagger}-\alpha^* \hat{a}\right)$ (which is the complex version of the Weyl operator) to a vacuum state.
The CF of a single-mode coherent state is given by
\begin{equation}\label{Eq:GeneralCoherentState}
    \chi_{\text{coh}}(\xi\,;\alpha)=\exp{\left[-\frac{|\xi|^2}{2}+\left(\xi \alpha^{*}-\xi^{*}\alpha\right)\right]},
\end{equation}
where $\xi$ can be treated as a complex variable.
A single-mode coherent state has mean value $\mathbf{\bar{x}}=[2\mathfrak{Re}(\alpha),2\mathfrak{Im}(\alpha)]^{\top}$ and a CM equals to $\mathbb{1}_2=\operatorname{diag}(1,1)$. 

A thermal state is a bosonic state which maximizes the von Neumann entropy  for fixed energy $\ensavg{\hat{a}^{\dagger}\hat{a}}=\bar{n}$, where $\bar{n}$ denotes the mean number of photons in the bosonic mode. The CF of a single-mode thermal state is given by
\begin{equation}
   \chi_{\text{thermal}}(\xi ;\bar{n})=\exp \left[-\left(2 \bar{n}+1\right)\frac{|\xi|^2}{2}\right],
\end{equation}
where $\xi$ can be treated as a complex variable. A single-mode thermal state has zero mean and CM $\mathbf{V}_{\text{thermal}}=(2\bar{n}+1)\mathbb{1}_2$.

A TMSV state $\ket{\psi_{\text{AB}}^{\text{TMSV}}}$ (with A and B denoting its two modes) is generated by applying the two-mode squeezing operator $S_t(r)=\exp \left[r\left(\hat{a}_{\text{A}} \hat{a}_{\text{B}}-\hat{a}_{\text{A}}^{\dagger} \hat{a}_{\text{B}}^{\dagger}\right) / 2\right]$ (with $r \in \mathbb{R}$ being the two-mode squeezing) to a pair of vacuum states $\ket{0}_{\text{A}}\ket{0}_{\text{B}}$. The level of two-mode squeezing is commonly quantified by $r_{\text{dB}}=-10 \log_{10}[\exp(-2r)]$ in dB. 
The CF of a TMSV state is given by
\begin{equation}
\begin{aligned}
\chi_{\operatorname{TMSV}}\left(\xi_{\mathrm{A}}, \xi_{\mathrm{B}};r\right)=  \exp \bigg\{-\frac{1}{2}\bigg[V_{\text{s}}\left(\left|\xi_{\mathrm{A}}\right|^2+\left|\xi_{\mathrm{B}}\right|^2\right)-\sqrt{V_{\text{s}}^2-1}\left(\xi_{\mathrm{A}} \xi_{\mathrm{B}}+\xi_{\mathrm{A}}^* \xi_{\mathrm{B}}^*\right)\bigg]\bigg\},
\end{aligned}
\end{equation}
where $\xi_\text{A}$ and $\xi_\text{B}$ can be treated as complex variables, and $V_{\text{s}}=\operatorname{cosh}(2r)$ is the quadrature variance of each mode.
A TMSV state has zero mean and the following CM
\begin{equation}\label{Eq:TMSV_CM}
\mathbf{V}_{\text{TMSV}}=\left[\begin{array}{cc}
V_{\text{s}} \mathbb{1}_2            & \sqrt{V_{\text{s}}^2-1} \mathbb{Z}_2 \\
\sqrt{V_{\text{s}}^2-1} \mathbb{Z}_2 & V_{\text{s}} \mathbb{1}_2
\end{array}\right],
\end{equation}
where $\mathbb{1}_2=\operatorname{diag}(1,1)$ and $\mathbb{Z}_2=\operatorname{diag}(1,-1)$.

\section{Usefulness of Uplink Configuration in Quantum Communications}\label{Sec:UplinkUsefulness}
In this section, we extend our discussion on the use of the uplink configuration for quantum communications and elaborate on its potential (and sometimes overlooked) usefulness within the context of current technological reach.

One may argue that the use of uplink channels should be entirely avoided since it is more challenging to transfer quantum signals in the uplink than in the downlink channels, and some of the well-known quantum information protocols can be implemented using solely downlink channels.
Indeed, the original GG02 CV-QKD protocol~\cite{GG02_Supp} can arguably be implemented without transferring any quantum signal in the uplink, and in its entanglement-based implementation, the QKD protocol can be carried out using the shared entanglement established in either uplink or downlink. 
However, it remains unclear whether the need for uplink channels can be completely eliminated when other quantum information protocols are used in a practical deployment. For example, many advanced QKD protocols, such as measurement-device-independent QKD (e.g.,~\cite{CV_MDI_QKD2015_Supp,SQCC_MDI_2020_Supp}), twin-field QKD (e.g., ~\cite{TFQKD2018_Supp,CV_TFQKD2019_Supp}), and more recently proposed protocols (e.g.,~\cite{Matt2021_Supp}), may naturally require the use of two uplink channels and a satellite to support the secret key generation between two distanced ground stations. 
For security reasons, using the uplink configuration also makes the attacks directed at receivers significantly more difficult in practice (see discussions in, e.g.,~\cite{UplinkBenefits_Supp}).

In addition, using the uplink configuration avoids deploying high-quality CV quantum entanglement sources on satellites, which is still impractical from a system complexity point of view.
Currently, most space-ready quantum entanglement sources (e.g.,~\cite{SpaceEntanglementSource2020_Supp}) generate polarization-entangled photon pairs, which are incompatible with CV quantum information protocols. 
A high-quality CV quantum entanglement source can take up considerable space in the payload and have stringent requirements such as ultra-stable laser light and fine-tuning of phase-matching achieved through precise adjustment and feedback control of the crystal temperature. 
The technical details revealing the high complexity of a high-quality squeezed laser source can be found in, e.g.,~\cite{SqueezedLaser2022_Supp} -- it should be noted that one standard method for generating strong and stable CV entanglement requires two squeezed laser sources like this (see, e.g.,~\cite{TMSV_Generation2013_Supp}). 
The on-chip generation of squeezing via nanophotonics represents one promising direction towards a space-ready CV quantum entanglement source; however, despite the recent advancements (e.g.,~\cite{5dB_SqueezingOnChip_Supp,10dB_SqueezingOnChip_Supp}), the on-chip generation of strong squeezing has yet to be demonstrated, let alone a space-ready CV entanglement source.
The use of the uplink configuration eliminates the need for (potentially complex) quantum entanglement sources in space, minimizing the complexity of the satellite and its payload. Such a minimization will be translated to reduced risk and cost as far as a practical deployment is concerned.

From an operational point of view, using the uplink configuration circumvents the significant limitations in flexibility and versatility of the downlink configuration.
Despite advancements in aerospace technologies, upgrading, reconfiguring, diagnosing, or even regularly maintaining a quantum light source (especially a CV entanglement source) remains a challenging task in the downlink configuration (where the source is onboard an in-orbit satellite). 
On the contrary, using the uplink configuration benefits from deploying quantum light sources at the ground stations, bringing great flexibility and allowing for future upgrades. 
The satellite payload can also be more versatile (than one containing the quantum light source) since it may naturally be compatible with various quantum information protocols, including those that simply require the satellite to play the role of a relay. 
In principle, a satellite-based system adopting the uplink configuration can support various quantum information protocols using different quantum light sources since the quantum light source can be readily reconfigured or switched out at the ground stations.

Despite the higher loss in the uplink configuration, it can be argued that deploying a high-quality CV quantum source is still much more challenging than fitting a sufficiently large telescope onto a satellite. 
In fact, the Micius experiment~\cite{UplinkTeleportation2017_Supp} has demonstrated that the satellite-borne telescope aperture for uplink quantum communications does not have to be very large. We also note that even for downlink quantum communications, the satellite still needs a sufficiently large telescope to ensure a good beam quality (and to reduce diffraction). Furthermore, it can be expected that the future adoption of advanced quantum information protocols (e.g.,~\cite{TFQKD2018_Supp,CV_TFQKD2019_Supp,Matt2021_Supp}) will allow for a higher loss tolerance, fully exploiting the advantages provided by the use of the uplink configuration.

\bigskip
\bigskip

\section*{Approved for Public Release: 24-1118}